\documentclass[11pt]{article}
\usepackage{graphicx}
\usepackage{natbib}

\usepackage{subfig}
\usepackage{amsmath}
\usepackage{amssymb}
\usepackage{txfonts}
\usepackage{color}
\usepackage[none]{hyphenat}

\usepackage{titlesec}

\usepackage{authblk}

\title{3-D Global modelling of the early martian climate under a dense CO$_2$+H$_2$ atmosphere and for a wide range of surface water inventories.}

\author[1,2]{Martin Turbet}
\author[2]{Francois Forget}

\affil[1]{Observatoire astronomique de l\'\ Universit\'e de Gen\`eve, Chemin Pegasi 51, 1290 Sauverny, Switzerland}
\affil[2]{Laboratoire de M\'et\'eorologie Dynamique/IPSL, CNRS, Sorbonne Universit\'e, Ecole Normale Sup\'erieure, PSL Research University, Ecole Polytechnique, 75005 Paris, France}

\date{\today}

\begin{document}

\maketitle

\begin{abstract}

CO$_2$+H$_2$ greenhouse warming has recently emerged as a promising scenario 
to sufficiently warm the early martian surface to allow the formation of valley networks and lakes. 
Here we present numerical 3-D global climate simulations of the early martian climate that we have performed assuming 
dense CO$_2$+H$_2$ atmospheres. Our climate model, derived from earlier works by \citet{Forget:2013} and \citet{Wordsworth:2013}, is coupled to an asynchronous model of the long-term evolution of martian glaciers and lakes. Simulations were carried out at 40$^\circ$ obliquity to investigate how (i) water content and (ii) H$_2$ content (added to 1 or 2 bars of CO$_2$) can shape the climate and hydrologic cycle of early Mars. We show that the adiabatic cooling mechanism \citep{Wordsworth:2013} that leads to the accumulation of ice deposits in the southern highlands in cold climate (the so called 'icy highland scenario') also works in warm climates, with impact crater lakes acting as the main water reservoirs. This produces rainfall mainly localized in the southern highlands of Mars. If one adjust (i) the amount of CO$_2$ and H$_2$, (ii) the size and location of the water reservoirs, and (iii) the ancient topography (i.e. by removing Tharsis), the spatial patterns of surface runoff (from rainfall or snowmelt) in the simulations can match -- with a few exceptions -- the observed distribution of valley networks and impact crater lakes. Although our results are obtained for CO$_2$-dominated atmospheres enriched with H$_2$, they should also apply to assess the impact of any combination of powerful long-lived greenhouse gases on early Mars.

\end{abstract}

\section{Introduction}
\label{section_introduction}

For several decades, planetary scientists have been trying to understand 
the origin of Late Noachian~/~early Hesperian Martian lakes and 
valley networks \citep{Carr:1995,Cabrol:1999,Malin:2003,Moore:2003,Mangold:2006,Hynek:2010} and the related mineralogical evidence for aqueous alteration forming, for instance, widespread hydrous clays  \citep{Poulet:2005,Carter:2015}. 
Climate modellers have explored a wide variety of mechanisms and effects including 
CO$_2$ greenhouse warming \citep{Pollack:1987,Halevy:2009,Wordsworth:2010,Ozak:2016,Turbet:2017jgr}, 
CO$_2$ ice clouds \citep{Forget:1997,Forget:2013,Kitzmann:2016}, 
H$_2$O ice clouds \citep{Urata:2013h2o,Wordsworth:2013,Ramirez:2017,Turbet:2020impact}, 
volcanic outgassing \citep{Tian:2010,Halevy:2014,Kerber:2015}, 
and extreme events such as outflow channel or chaos-forming events \citep{Kite:2011b,Turbet:2017icarus} 
and bolide impacts \citep{Segura:2002,Segura:2008,Segura:2012,Turbet:2018,Steakley:2019,Turbet:2020impact}.
However, none of these scenarios has so far given satisfactory results, i.e. that could 
reasonably explain the formation of Martian lakes and river beds.

While reducing gases (e.g. CH$_4$, NH$_3$, H$_2$) have been proposed by \citet{Sagan:1977} as a possible 
source of greenhouse warming in the atmosphere of early Mars more than fourty years ago, 
true interest in this possibility has only really been considered recently \citep{Ramirez:2014,Wordsworth:2017}. This stems from the fact that prior to the works of \citet{Ramirez:2014} and \citet{Wordsworth:2017}, this scenario was not very compelling for the following reasons:
\begin{itemize}
\item The calculated levels of di-hydrogen (H$_2$) required to warm the surface of early Mars in a CO$_2$-dominated 
atmosphere were so high that a very strong source of production of H$_2$ was required to offset 
the efficient H$_2$ atmospheric escape to space \citep{Ramirez:2014}.
\item The calculated levels of methane (CH$_4$) required to warm the surface of early Mars in a CO$_2$-dominated 
atmosphere were so high that CH$_4$ should have been photodissociated by incoming XUV solar photons, forming highly 
reflective (and thus radiatively cooling) photochemical hazes. Moreover, stratospheric CH$_4$ can absorb 
a significant fraction of the incoming solar radiation and thus can produce an 
anti-greenhouse effect \citep{Haqq-misra:2008,Wordsworth:2017}.
\item Ammonia (NH$_3$) should have been efficiently photodissociated by XUV solar photons, requiring an unrealistically strong source. H$_2$ would have 
escaped to space and the remaining N$_2$ should have built up in the atmosphere. Yet, we do not see any evidence 
supporting this scenario in present-day Mars atmosphere.
\end{itemize}
Recently, several works showed that collision-induced absorptions (CIAs) between H$_2$/CH$_4$ and CO$_2$ molecules can in fact provide a very strong source of greenhouse warming on early Mars \citep{Ramirez:2014,Wordsworth:2017,Turbet:2019spectro,Ramirez:2019rnaas,Turbet:2020spectro,Godin:2020,Hayworth:2020,Mondelain:2021,Wordsworth:2021}. This relaxes the constraint on the amount and the production rate of H$_2$ -- of different possible origins, including but not limited to reducing volcanism \citep{Ramirez:2014,Liggins:2020}, serpentinization \citep{Chassefiere:2016}, radiolysis \citep{Tarnas:2018}, impacts \citep{Haberle:2019} -- needed to warm the surface of early Mars, and thus gives credibility to this hypothesis \citep{Wordsworth:2021}. A comprehensive description of the incremental contributions of these works is provided in the Introduction Section of \citet{Turbet:2020spectro}. In summary, \citet{Ramirez:2014}, \citet{Wordsworth:2017} and \citet{Turbet:2020spectro} used 1-D climate models to compute the amount of H$_2$ required to warm the surface of early Mars above the melting point of water (see Table~\ref{table_cia_tsurf}). Their results mainly differ in the way they took into account the effect of CO$_2$+H$_2$ CIAs. \citet{Ramirez:2014} used the CIA of N$_2$+H$_2$ instead of that of CO$_2$+H$_2$ due to the lack of relevant data. \citet{Wordsworth:2017} evaluated the CO$_2$+H$_2$ CIA based on ab-initio calculations and empirical scalings. \citet{Turbet:2020spectro} carried out experimental measurements of the CO$_2$-H$_2$ CIA (based on the earlier work of \citealt{Turbet:2019spectro}) which they then used to build a semi-empirical model of the CO$_2$-H$_2$ CIA. To date, this is arguably the most reliable database for the collision-induced absorption of CO$_2$+H$_2$.

\begin{table*}
\centering
\begin{tabular}{lcl}
   Amount of CO$_2$ & 1~bar & 2~bar \\
   \hline
   \citet{Ramirez:2014} & \textgreater 20$\%$ & $\sim$~15$\%$ \\
   \citet{Wordsworth:2017} & $\sim$~10$\%$ & $\sim$~2.5$\%$ \\
   \citet{Turbet:2020spectro}~/~this work & $\sim$~27$\%$ & $\sim$~7$\%$ \\
   \hline
\end{tabular}
\caption{H$_2$ content (in $\%$) required to warm early Mars above 273K, depending on the assumed CO$_2$ partial pressure (1 or 2~bar). These results were obtained with 1-D numerical climate models, 
and using different estimates of the CO$_2$-H$_2$ collision-induced absorptions (see main text). The H$_2$ content (volume mixing ratio) is defined as $\frac{\text{n}_{\text{H2}}}{\text{n}_{\text{CO2}}+\text{n}_{\text{H2}}}$, with n$_\text{X}$ being the number of moles of molecule X.}
\label{table_cia_tsurf}
\end{table*}

\medskip

The warming of early Mars by the combined action of CO$_2$ and H$_2$ has been studied mainly with 1-dimensional climate models \citep{Ramirez:2014,Wordsworth:2017,Turbet:2020spectro}. These models cannot take into account key components of the climate system (atmospheric dynamics, clouds, effect of topography, continental hydrology, etc.). To the best of our knowledge, only three studies have explored the climate of early Mars using 3-dimensional Global Climate Models (GCM) and assuming a greenhouse effect sufficiently strong to create a warm climate. These three studies have investigated different aspects of how a significant warming of the early Martian atmosphere may affect the hydrologic cycle and thus the precipitation patterns on the Martian surface. First, \citet{Wordsworth:2015} used the 3-D "LMD Generic Early Mars" GCM to simulate strong greenhouse warming under the idealized grey gas approximation on early Mars. 
Secondly, \citet{Turbet:2019NatSR} used the same GCM taking into account absorption by CO$_2$, H$_2$ and H$_2$O to explore the stability of a Late Hesperian ocean under a warm, CO$_2$+H$_2$-dominated atmosphere. 
Thirdly, \citet{Kamada:2020} used the 3-D "DRAMATIC" GCM to explore how the addition of hydrogen to the early Martian CO$_2$-dominated atmosphere impacts the hydrologic cycle. Both \citet{Turbet:2019NatSR} and \citet{Kamada:2020} used the CO$_2$+H$_2$ CIA tables provided in \citet{Wordsworth:2017}. 
The most important findings of these studies in the context of our work is that (1) CO$_2$+H$_2$ greenhouse warming is also very efficient in 3-D numerical climate models to warm the surface and atmosphere of Mars above the melting point of water, and can thus trigger an intense water cycle \citep{Turbet:2019NatSR,Kamada:2020}; and that (2) warm and wet climates triggered by a strong greenhouse warming can produce precipitation (possibly rainfall) but that these precipitations do not appear to be spatially correlated with the known locations of valley networks \citep{Wordsworth:2015,Kamada:2020}. These two latter studies will be extensively discussed and compared over the course of this manuscript.

Our present work is somehow an extension of \citet{Wordsworth:2015} who focused only on the case of large water content scenarios (large enough to cover a significant fraction of Mars by open water oceans) and used a highly idealized description of greenhouse warming following the grey gas approximation. Here we explore how the climate of early Mars would look like depending on how much CO$_2$, H$_2$ and H$_2$O are present in the atmosphere and/or on the surface of the planet. Compared to previous works, several significant advances have been made in our work. Firstly, we used the most recent CO$_2$+H$_2$ CIA look-up tables of \citet{Turbet:2020spectro} to evaluate the greenhouse warming by CO$_2$+H$_2$-rich atmospheres. Secondly, we took into account a realistic model of paleo-topography of Mars -- at the time most valley networks formed, near the Late Noachian era -- based on the work of \citet{Bouley:2016}. Thirdly, we adapted the 3-D "LMD Generic Early Mars" GCM simulations to account for the effect of oceans (considering several sizes) and more importantly for the long-term effect of impact crater lakes that would form in a warm climate. For this purpose the GCM is coupled to a simple asynchronous model of the long-term evolution of possible martian glaciers and lakes.

The manuscript is organized as follows. In Section~\ref{reducing_gas_warming_method}, we present the early Mars version of the 3-D LMD Generic Global Climate Model, with a 
focus on all the improvements made (1) to account for the greenhouse warming produced by reducing gases 
and (2) to account for hydrologic feedbacks (oceans, lakes and glaciers).
In Section~\ref{reducing_section_amount_hydrogen}, we reassess the amount of hydrogen required to produce 
fluvial activity on Mars, based on the results of 3-D climate simulations. We then explore in Section~\ref{reducing_section_warm_climates} the nature(s) of the early Mars climate and hydrology for strong greenhouse warming scenarios, and for a wide range of possible total water inventories. Eventually, discussions and conclusions are provided in Section~\ref{conclusions_discussions}.

\section{Method}
\label{reducing_gas_warming_method}

We used the LMD Generic Model, a full 3-Dimensions Global Climate Model (GCM) that initially derives 
from the LMDz Earth \citep{Hourdin:2006} and Mars \citep{Forget:1999} GCMs. 
This GCM has previously been developed and used for the study of the climate of ancient Mars 
\citep{Forget:2013,Wordsworth:2013,Wordsworth:2015,Kerber:2015,Turbet:2017icarus,Palumbo:2018,Turbet:2019NatSR,Turbet:2020impact}.

\subsection{General description of the model}

Simulations presented in this paper were performed at a horizontal resolution of 64~$\times$~48 (e.g. 5.6$^{\circ}$~$\times$~3.75$^{\circ}$; 
at the equator, this gives in average 330~km~$\times$~220~km) in longitude~$\times$~latitude. 
In the vertical direction, the model is composed of 26 distinct atmospheric layers, covering altitudes 
from the surface up to $\sim$~10~Pascals. Hybrid $\sigma$ coordinates (where $\sigma$ is the ratio between pressure and surface pressure) 
and fixed pressure levels were used in the lower and the upper atmosphere, respectively.
The dynamical time step of the simulations is 92~s. The radiative transfer and the physical parameterizations are calculated every $\sim$~1~hour and  $\sim$~15~minutes, respectively. 
Parameterizations of turbulence and convection, full CO$_2$ and H$_2$O cycles (condensation, evaporation,
sublimation, cloud formation, precipitation, etc.), were all parameterized as in \citet{Wordsworth:2013} and \citet{Turbet:2017icarus}. 
We focus below on the new parameterizations developed specifically for the present study.

\begin{table*}
\centering
\begin{tabular}{ll}
   \hline
              &              \\
   Physical parameters & Values \\
              &              \\
   \hline
              &              \\
   Mean Solar Flux & 111~W~m$^{-2}$  (75$\%$ of present-day Mars) \\
   Bare ground albedo & 0.2 \\
   Liquid water albedo  & 0.07 \\
   H$_2$O ice albedo  & 0.55 \\
   CO$_2$ ice albedo  & 0.5 \\
   Obliquity & 40$^\circ$ \   \citep[$\sim$ most probable value,][]{Laskar:2004}\\
   Orbital eccentricity & 0 \ (circular orbit) \\
   Surface Topography & Pre-True Polar Wander (modified if oceans are present) \\
   Surface roughness coefficient & 0.01 m \\
   Ice thermal inertia & 1500 J~m$^{-2}$~s$^{-1/2}$~K$^{-1}$ \\
   Ground thermal inertia & 250+7~x$_{\text{H}_2\text{O}}$. J~m$^{-2}$~s$^{-1/2}$~K$^{-1}$, where x$_{\text{H}_2\text{O}}$ is the soil moisture (in kg~m$^{-3}$)  \\
   CO$_2$ partial pressure & 1 or 2 bars \\
   No. of CO$_2$ ice cloud cond. nuclei (CCN) & $10^5$~kg$^{-1}$ \ \citep{Forget:2013} \\
   No. of H$_2$O cloud cond. nuclei (CCN) & $4\times10^6$~kg$^{-1}$ (liquid), $2\times10^4$~kg$^{-1}$ (ice) \ \citep{Leconte:2013nat}\\
   H$_2$O precipitation threshold & 0.001 kg kg$^{-1}$ \\
   
              &              \\
   \hline

\end{tabular}
\caption{Summary of the main physical parameters used in the GCM for this study.}
\end{table*}

\subsubsection{Radiative Transfer}

The GCM includes a generalized radiative transfer code able to account for a variable 
gaseous atmospheric composition made of a mixture of CO$_{2}$, H$_2$, and H$_{2}$O (line list and 
parameters were taken from HITRAN 2008 database \citep{Rothman:2009}) using the 'correlated-k' approach \citep{Fu:1992} suited 
for fast calculation. Our correlated-k absorption coefficients directly derive 
from \citet{Wordsworth:2013}. They take into account the CO$_2$ Collision-Induced Absorptions \citep{Gruszka:1998,Baranov:2004,Wordsworth:2010}, 
as well as the H$_2$O MT$\_$CKD continua \citep{Clough:2005}.
Compared to \citet{Wordsworth:2013}, we added the radiative effect of H$_2$ through self H$_2$-H$_2$ and foreign 
CO$_2$-H$_2$ collision-induced absorptions (CIA). H$_2$-H$_2$ CIAs were taken from HITRAN CIA database \citep{Richard:2012,Karman:2019}.
CO$_2$-H$_2$ CIAs were taken from \citet{Turbet:2020spectro}.

We adopted here a mean solar flux of 111~W.m$^{-2}$ (75$\%$ of the present-day 
value of Mars; 32$\%$ of Earth's present-day value; as in \citealt{Wordsworth:2013}), 
corresponding to the reduced luminosity derived from standard solar evolution models \citep{Gough:1981} 3.7~Gya, 
during the Late Noachian era.

\subsubsection{Topography}

\begin{figure}
\centering
\includegraphics[scale=0.2]{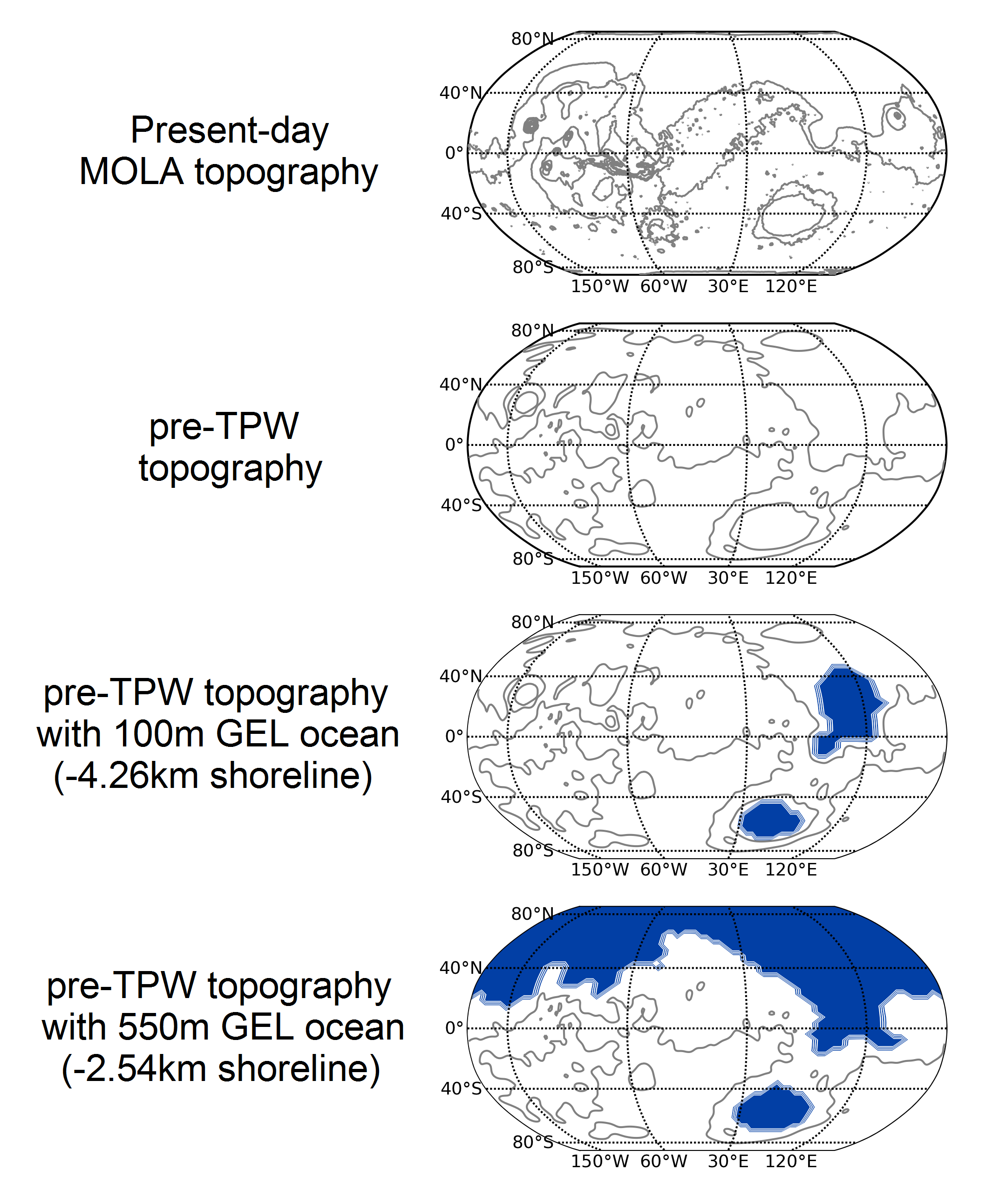}
\caption{Topography maps discussed and used in our work. 
The topography map (top row) is the present-day Mars MOLA topography \citep{Smith:1999,Smith:2001}. 
Although it is not used in our study, we put it here as a reference.
The second row corresponds to the pre-TPW (pre-True Polar Wander) topography map \citep{Bouley:2016}. The third and fourth rows 
correspond to the pre-TPW topography maps with the inclusion of ancient oceans (with -4.26 and -2.54~km shorelines, respectively). }
\label{figure_reducing_topo_earlymars}%
\end{figure}

We used the pre-True Polar Wander (pre-TPW) topography (see Fig~\ref{figure_reducing_topo_earlymars}, second row) 
from \citet{Bouley:2016}. \citet{Bouley:2016} showed that the formation of the late Noachian 
valley networks is likely to have predated the formation of most of the Tharsis volcanic bulge. 
The pre-TPW topography is based on the present-day MOLA (Mars Orbiter Laser Altimeter) Mars surface 
topography shown in Figure~\ref{figure_reducing_topo_earlymars}, first row \citep{Smith:1999,Smith:2001}, 
but without Tharsis and all the younger volcanic features. 
Moreover, the formation of Tharsis has been shown to have produced a large True Polar Wander (TPW) event of 20$^{\circ}$-25$^{\circ}$, 
which is also taken into account in the pre-TPW topography. This topography was adopted for numerical climate 
simulations assuming low water contents (low enough that no permanent ocean is formed).
In the simulations that account for an ancient ocean, the 
pre-TPW topography was adjusted (see Fig~\ref{figure_reducing_topo_earlymars}, third and fourth rows) 
so that the minimum altitude were -4.26 and -2.54~km\footnote{These negative altitudes are defined according 
to the reference MOLA topography map \citep{Smith:1999,Smith:2001}.}:
\begin{itemize}
\item The -4.26~km shoreline case (third row) was calculated assuming a total surface water inventory of $\sim$~100~m GEL.
\item The -2.54~km shoreline case (fourth row) corresponds to the putative northern ocean 
shoreline based on delta deposit locations from \citet{Diachille:2010}. $\sim$~550~m of global equivalent layer (GEL)
is needed to reach this shoreline. \citet{Wordsworth:2015} and \citet{Kamada:2020} adopted a similar ocean 
elevation, but used the present-day Mars MOLA topography.
\end{itemize}

All regions matching these elevations were then treated as 'oceanic regions'.

\subsubsection{Parameterization of oceanic regions}

We used the simplified ocean model from \citet{Codron:2012} to treat oceanic regions. This model has been 
previously used to explore the past climate(s) of the Earth \citep{Charnay:2013}. The modeled ocean is composed 
of two layers. The first upper layer (50~m deep) represents the surface mixed layer, where the exchanges with the atmosphere take place. 
The second lower layer (150~m deep) represents the deep ocean. 
The oceanic model computes the formation of sea ice. Sea ice forms when the ocean temperature
falls below 273~K and melts when its temperature rises above 273~K. The changes in ice extent and thickness are 
computed based on energy conservation, keeping the ocean temperature at 273~K as long as ice is present. 
A layer of snow can be present above the ice, and the surface albedo is calculated accordingly \citep{Codron:2012,Charnay:2013}.
The transport of heat by the ocean circulation is not taken into account here (nb: it is provided by the model, but not yet available in a parallelized form). The transport of sea ice is not either taken into account.

\subsubsection{Parameterization of continental regions}

\label{sc:continental}

\paragraph{Soil Model}

\begin{figure*}
\centering
\includegraphics[width=12cm]{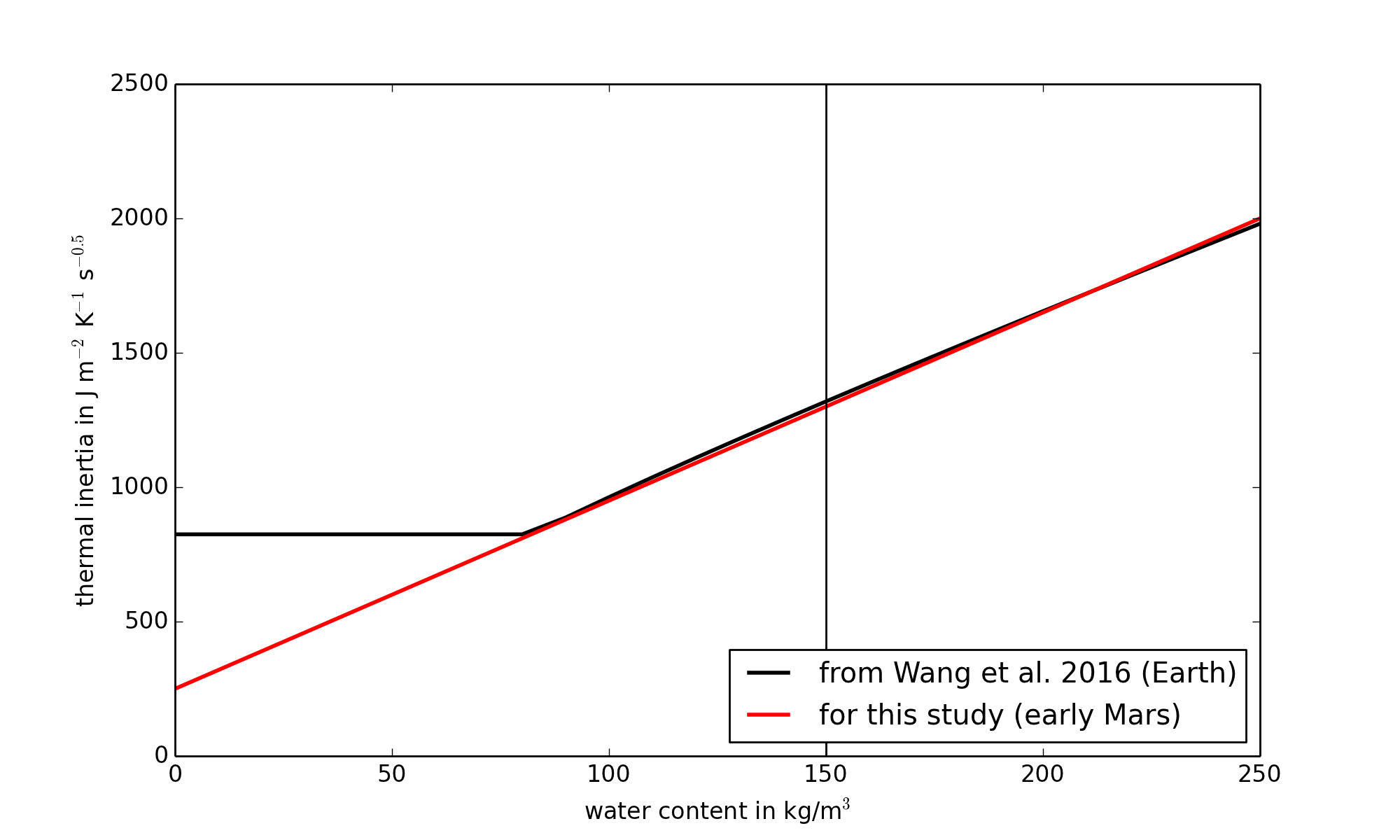}
\caption{Thermal inertia of the soil as a function of the liquid water loading (in kg/m$^3$). The black line corresponds to the standard parameterization used in the ORCHIDEE Earth land model \citep{Wang:2016}. The red curve corresponds to the parameterization that we used in the present study to account for the thermal inertia of the Martian ground in early Mars conditions. Our parameterization is compatible with both (1) a standard dry ground thermal inertia of 250~J~m$^{-2}$~s$^{-1/2}$~K$^{-1}$ and (2) the water-loading-dependency of the thermal inertia used in the ORCHIDEE Earth land model.}
\label{figure_reducing_thermal_inertia}
\end{figure*}

To account for the thermal conduction in the subsurface in continental regions (i.e. all non-oceanic regions), 
we used a 20-layers thermal diffusion soil model. The mid-layer depths 
range from d$_0$~$\sim$~0.15~mm to d$_{19}$~$\sim$~80~m, following the power law 
d$_n$~=~d$_0$~$\times$~2$^n$ with $n$ being the corresponding soil level, chosen to take 
into account both the diurnal and seasonal thermal waves.

We assumed the thermal inertia of the regolith I$_{\text{ground}}$ to be equal to:
\begin{equation}
    I_{\text{ground}}=I_{\text{dry}}+7~x_{\text{H}_2\text{O}}, 
\end{equation}
where I$_\text{dry}$~=~250~J~m$^{-2}$~s$^{-1/2}$~K$^{-1}$ and x$_{\text{H}_2\text{O}}$ is the soil moisture (in kg~m$^{-3}$). 
The dry regolith thermal inertia is slightly higher than the present-day Mars global mean 
thermal inertia in order to account for the higher atmospheric pressure \citep{Piqueux:2009} which increases the thermal conductivity of the soil because of the gas in the pores of the soil. 
This expression (plotted in Fig~\ref{figure_reducing_thermal_inertia}) has been derived from 
the standard parameterization of the ORCHIDEE (Organising Carbon and Hydrology In Dynamic Ecosystems) 
Earth land model \citep{Wang:2016}.
Moreover, we arbitrarily fixed the thermal inertia of the ground 
to a value of 1500~J~m$^{-2}$~s$^{-1/2}$~K$^{-1}$, whenever the snow/ice cover exceeds a threshold of 1000~kg~m$^{-2}$ (e.g. $\sim$~1m).

We assumed that the Martian regolith has a maximum water capacity of 150~kg~m$^{-2}$, 
based on a simple bucket model widely used in the Earth land community \citep{Manabe:1969,Wood:1992,Schaake:1996}
\footnote{This is actually known as the "Manabe" bucket model.}. 
When the water quantity exceeds this limit, the overage is treated as runoff. 
Similarly, we limit the maximum amount of snow/ice surface deposits to 3000~kg~m$^{-2}$ (i.e. $\sim$~3~m).

\paragraph{Parameterization of ancient impact crater lakes in the GCM}

The continental or unsubmerged regions of the planet are covered by a large number of 
craters that can potentially host lakes. It is crucial to include these lakes in early Mars 
climate models because the presence of lakes controls whether the different regions of Mars are 
desertic or not, which may have further effects on regions of precipitation. The accumulation of water in impact crater lakes can change the intensity and spatial distribution of evaporation and, 
more broadly, affect the global hydrologic cycle.

For each unsubmerged region (i.e. non-oceanic region), 
we allowed a fraction $\alpha_\text{lake}$ of up to 0.49 (i.e. 49$\%$) of the surface to be covered by lakes. 
According to the model of \cite{Matsubara:2011}, this is the maximum possible fraction of submerged regions in the highly craterized Martian terrains, before all the lakes overflow .

In our parameterization, the contribution of lakes is twofold: 
\begin{itemize}
\item First, we consider them as a source of water vapor, 
and we calculate their combined rate of evaporation $E$  (or sublimation, if the temperature is below 273~K) averaged over the grid mesh (in kg~m$^{-2}$~s$^{-1}$) as follows:
\begin{equation}
E~=~\alpha_{\text{lake}}~\rho_{1} C_d V_{1} [q_{\text{sat}}(T_{\text{surf}})-q_{1}] 
\label{evap}
\end{equation}
where $\rho_{1}$ and V$_{1}$ are the volumetric mass of air (kg~m$^{-3}$) and the wind velocity (m~s$^{-1}$) in the first atmospheric model level, 
$q_{\text{sat}}(T_{\text{surf}})$ is the water vapor mass mixing ratio at saturation at the surface (kg~kg$^{-1}$), 
and $q_{1}$ is the mass mixing ratio in the first atmospheric layer. The aerodynamic coefficient is given by 
\begin{equation}
C_d~=~{(\kappa / ln(1+z_1 / z_0))}^2~\sim~2.5\times10^{-3}
\label{cd}
\end{equation}
where $\kappa$~=~0.4 is the Von Karman constant, $z_0$ is the surface roughness coefficient (set to 0.01~m) and 
$z_1$ is the altitude of the first atmospheric level ($\sim$~20~meters).
Simultaneously, evaporation from the surrounding regolith terrains (when wet or ice-covered, for instance after precipitation) is then multiplied by a factor (1-$\alpha_\text{lake}$).
\item Secondly, we took into account the effect of lakes on the surface albedo. 
If the surface temperature is below or equal to 273~K, the albedo of the lakes is 0.55; otherwise it is 0.07.
\end{itemize}

The formation and evolution of the impact crater lakes are discussed in the next subsection.

\subsection{Determining where glaciers and lakes stabilize}

All GCM simulations were first run for 10~Martian years as in \citealt{Forget:2013} to reach a first 
atmospheric equilibrium, with surface and atmospheric temperatures roughly equilibrated. 
Yet, it would take several orders of magnitude longer for the water cycle to reach an equilibrium. The reason is twofold:
\begin{itemize}
\item First, in the regions that are cold enough for permanent ice deposits to exist, 
sublimation and light snowfall are the dominant forms of water transport. Water ice distribution 
can then take thousands of years and even much more to reach a steady state \citep{Wordsworth:2013,Turbet:2017icarus}. 
\item Secondly, the evolution of impact crater lakes to their final size and location can also take a very long time, 
because this evolution results from a slow competition between evaporation of the lake, 
precipitation on the lake, and runoff from surrounding terrains.
\end{itemize}

\subsubsection{Evolution of ice deposits}

The locations of stable surface ice deposits were calculated using the ice equilibration algorithm of \citet{Wordsworth:2013}, 
later used in \citet{Turbet:2017icarus}. After the first ten Martian years of evolution, we proceeded with the following repeated steps: 
\begin{enumerate}
 \item We run the GCM for two Martian years.
 \item We extrapolate the evolution of the ice layer field $h_{\text{ice}}$ over $n_{\text{years}}$ number of years using:
\begin{equation}
h_{\text{ice}}(t+n_{\text{years}})=h_{\text{ice}}(t)+n_{\text{years}}~\times~\Delta h_{\text{ice}},
\label{ice_iteration_scheme}
\end{equation}
with $\Delta h_{\text{ice}}$ the annual mean ice field change of the one-Martian-year previous simulation.
\item We eliminate the seasonal ice deposits.
\item We limit the extrapolated ice field to a maximum value of 3000~kg~m$^{-2}$ (i.e. $\sim$~3~m).
\item We repeat the process.
\end{enumerate}

This algorithm was shown \citep{Wordsworth:2013} to be insensitive to the initial ice field location at the beginning of the simulation, 
assuming that the scheme was repeated a sufficient number of times. In total, for our control simulations, we looped the algorithm 20 times, with $n_{\text{years}}$=50~years for the first 10 loops 
and $n_{\text{years}}$=20 for 10 more loops. Because our simulations are much warmer than those of \citet{Wordsworth:2013}, 
the water transport is much more efficient, and the convergence of the algorithm is more rapid.
Note that, for the cases of "dry" simulations (those that do not include oceans), we arbitrarily normalized the total amount of permanent water ice deposits to 1~m GEL after each iteration of the ice equilibrium algorithm.

\subsubsection{Formation and evolution of impact crater lakes}

\begin{figure*}
\centering
\includegraphics[width=\columnwidth]{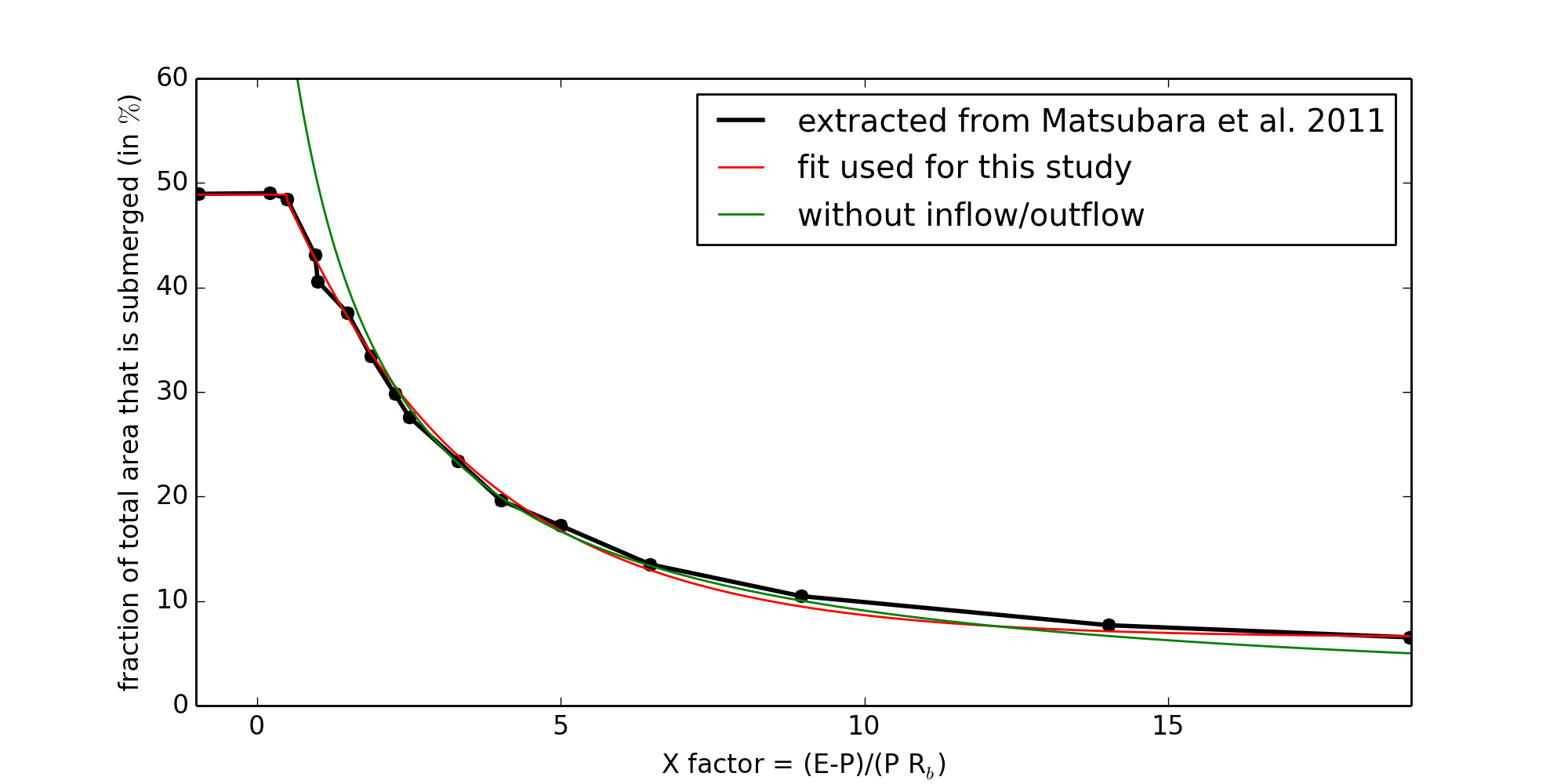}
\caption{Fraction of lakes area (black line) covering a craterized terrain as a function of the aridity index ``X-ratio'' (see text), derived from the results of \citet{Matsubara:2011}. The red line is the fit that we used for the lake evolution algorithm. Without inflow$/$outflow, $\alpha_{\text{lake}}$~=~$\frac{1}{X+1}$, corresponding to the green line.}
\label{fig_reducing_fraction_submerged}
\end{figure*}

To calculate the formation and evolution of impact crater lakes, we 
developed an algorithm based on the results of the Martian hydrology model of \citet{Howard:2007} and \citet{Matsubara:2011}. 
Our algorithm relies on the ``X-ratio'' (a proxy for the aridity of the region), defined as: 
\begin{equation}
    X=\frac{E-P}{P R_{\text{b}}},
\end{equation}
with $P$ the annual mean rate of precipitation (m/year), $E$ the annual mean rate of evaporation above the lakes only 
(m/year), $R_{\text{b}}$ is the fraction of precipitation above the unsubmerged land and that ends up in runoff, 
in the lake. These three quantities can be recorded at each grid point in our Global Climate Model simulations. \citet{Matsubara:2011} provides a useful relationship between the surface fraction of lakes $\alpha_{\text{lake}}$ with this aridity index X-ratio, 
for a highly craterized terrain typical of Late Noachian terrains. 
This relationship is plotted in Fig~\ref{fig_reducing_fraction_submerged} (black line). 
We fit this relationship (red line) as follows: 
\begin{equation}
    \begin{cases}
    \alpha_{\text{lake}}~=~49~\text{for}~X<0.5 \\
    \alpha_{\text{lake}}~=~6.5+42.5~e^{\frac{0.5-X}{3.2}}~\text{for}~0.5<X<19 \\
    \alpha_{\text{lake}}~=~0~\text{for}~X>19
    \end{cases}
\label{eq_x_factor}
\end{equation}
where $\alpha_{\text{lake}}(X)$ is expressed in $\%$. 
In short, a low value of X means that the lakes must grow in coverage to increase the evaporation surface, 
and thus reach hydrological equilibrium. A high value of X, on the other hand, indicates that lakes must reduce in size 
to limit the rate of evaporation and thus also reach hydrological equilibrium.

Our lake formation and evolution algorithm (run in parallel with the ice equilibrium algorithm described above) is as follows: After the first ten Martian years of evolution we proceed with the following steps: 
\begin{enumerate}
 \item We run the GCM for two Martian years. During the second year, at each timestep we record (i) the precipitation over each grid mesh, 
(ii) the evaporation of the lakes over each grid mesh (=~0 if there is no lakes) and (iii) the runoff from unsubmerged lands to lakes. In each grid mesh, we assume that 100$\%$ of the water that runoff (when the soil water content exceeds 150~kg~m$^{-2}$) end up in the lakes of the same grid mesh.
\item We calculate the aridity X-ratio for each GCM grid. 
\item We estimate $\alpha_{\text{lake}}(X)$ based on equations~\ref{eq_x_factor}. We limit the variation of the coverage of 
the lakes to 10~$\%$ from one step of the algorithm to the next, in order to avoid numerical unstabilities.
\end{enumerate}

Note that the last step of this algorithm does not conserve water. This is an issue only for the dry and warm simulations 
(those for which the entire water inventory can be trapped in impact crater lakes). For the dry 
(dry enough so that no permanent ocean is present) and warm (warm enough so that water is not fully trapped as ice) simulations, we arbitrarily normalize the total amount of water stored in impact crater lakes to the initial total water content (e.g. arbitrarily fixed to 4~m GEL here, for the dry simulations) 
after each iteration of the lake evolution algorithm. We discuss the impact of this renormalization (with the help of sensitivity studies) in Section~\ref{reducing_section_warm_climates}. 

\subsection{Initial states of the simulations}

In this section we describe the initial simulation setups, summarized in Table~\ref{reducing_experiments}. 
Initially, the surface/subsurface/atmosphere 
temperatures were arbitrarily fixed to 300~K everywhere on the planet.

We distinguish two main categories of initial states, depending on the total water content:
\begin{itemize}
\item \textbf{Cases where early Mars is water-rich (i.e. oceans are present)}. In these scenarios, 
we initialized non-oceanic 
regions with completely dry conditions. Oceanic regions were assumed to be initially fully deglaciated, with a uniform temperature of 300~K.
\item \textbf{Cases where early Mars is water-poor (i.e. no oceans are present)}. In these scenarios, we assumed that all the water was initially trapped as liquid water in impact crater lakes uniformly distributed on the surface. The initial total water content is arbitrarily fixed to 4~m GEL. For the warm simulations (warm enough so that water is not fully trapped as ice), the lake algorithm renormalization step maintains the total amount of water at 4~m GEL. For the cold simulations (cold enough that water can accumulate as ice on the coldest points of the planet), the lake algorithm renormalization step is not performed, and the amount of water is therefore not conserved. In this particular case, the amount of water is highly dependent on the amount of water trapped as ice. The final amount of water is not strictly speaking fixed because it depends on the size of the glaciers considered.
\end{itemize}

\begin{figure*}
\centering
\includegraphics[width=12cm]{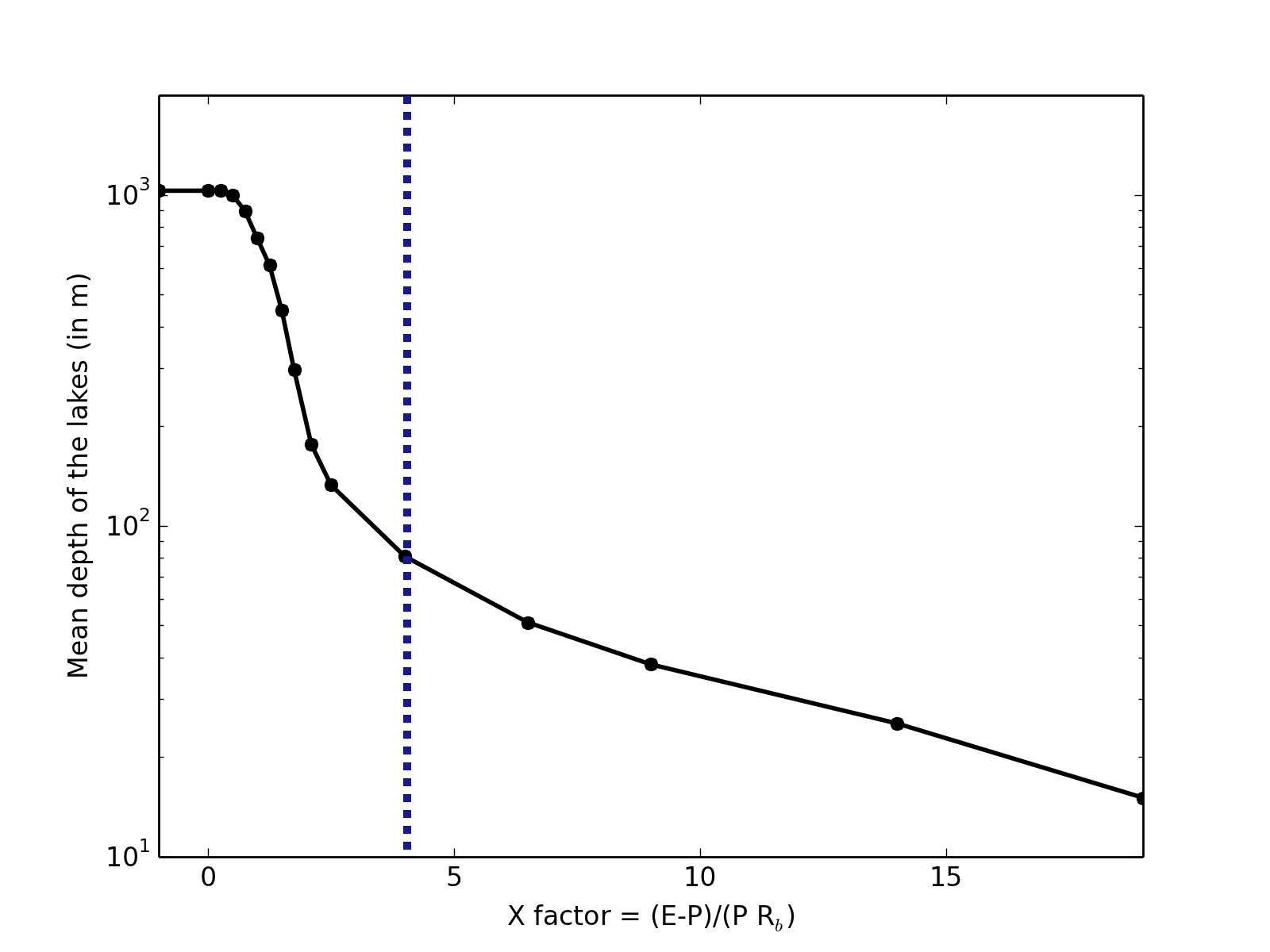}
\caption{Mean lake depth (in m) versus aridity X-ratio, derived from the results of \citet{Matsubara:2011}. Note that the mean lake depth at low X-ratio (typically lower than 4, denoted by the position of the dotted, vertical line) is biased by the contribution of the largest impact craters (Hellas, Argyre, etc.).}
\label{fig_reducing_lake_depth}
\end{figure*}

The amount of water h$_{\text{H}_2\text{O}}$ trapped in impact crater lakes in each grid mesh can be calculated using 
Figs.~\ref{fig_reducing_fraction_submerged} and \ref{fig_reducing_lake_depth}, based on \citet{Matsubara:2011}. Using a hydrologic routing model coupled with Mars topography, \citet{Matsubara:2011} estimated the averaged relationship between the surface area of lakes and their averaged depth on the craterized ancient surface of Mars. We fitted these result with the following empirical relationship (that excludes the contribution of the largest Martian craters):
\begin{equation}
h_{\text{H}_2\text{O}}~=~21.7-\frac{26}{(1+\alpha_\text{lake}/100)^4}
\label{equation_GEL_versus_coverage}
\end{equation}
with $h_{\text{H}_2\text{O}}$ the amount of water in lakes averaged over the grid mesh area (equivalent layer in meters, or 1000~kg m$^{-2}$) and $\alpha_\text{lake}$ in $\%$. $h_{\text{H}_2\text{O}}$. $h_{\text{H}_2\text{O}}$ can be integrated over the planet to provide the total inventory of water trapped in lakes.
For reference, the total amount of water that can be trapped in impact craters before all lakes overflow 
is $\sim$~60~m and 500~m GEL, without and with the contribution of the largest Martian craters (mainly Hellas and Argyre), respectively.

The list of simulations performed and discussed in our work is summarized in Table~\ref{reducing_experiments}.

\begin{table*}
\centering
\caption{List of simulations discussed in this work. Additional simulations (not listed here) were also conducted for sensitivity experiments (in particular, relative to the lake algorithm). These sensitivity experiments are discussed later in the manuscript. The amount of additional H$_2$ is reported in volume mixing ratio.}
\begin{tabular}{lccccc}
\hline
Simulation & CO$_2$ surface pressure & Amount of additional H$_2$ & Ocean \\
\hline
1 & $1$~bar & 5$\%$ & yes (550m GEL) & \\
2 & $1$~bar & 10$\%$ & yes (550m GEL) & \\
3 & $1$~bar & 15$\%$ & yes (550m GEL) & \\
4 & $1$~bar & 18$\%$ & yes (550m GEL) & \\
5 & $1$~bar & 22$\%$ & yes (550m GEL) & \\
6 & $1$~bar & 30$\%$ & yes (550m GEL) & \\

7 & $1$~bar & 5$\%$ & no & \\
8 & $1$~bar & 10$\%$ & no & \\
9 & $1$~bar & 15$\%$ & no & \\
10 & $1$~bar & 18$\%$ & no & \\
11 & $1$~bar & 22$\%$ & no & \\
12 & $1$~bar & 30$\%$ & no & \\

13 & $2$~bar & 0.5$\%$ & yes (550m GEL) & \\
14 & $2$~bar & 1$\%$ & yes (550m GEL) & \\
15 & $2$~bar & 2$\%$ & yes (550m GEL) & \\
16 & $2$~bar & 3$\%$ & yes (550m GEL) & \\
17 & $2$~bar & 4$\%$ & yes (550m GEL) & \\
18 & $2$~bar & 6$\%$ & yes (550m GEL) & \\
19 & $2$~bar & 8$\%$ & yes (550m GEL) & \\
20 & $2$~bar & 10$\%$ & yes (550m GEL) & \\
21 & $2$~bar & 15$\%$ & yes (550m GEL) & \\
22 & $2$~bar & 1$\%$ & no & \\
23 & $2$~bar & 2$\%$ & no & \\
24 & $2$~bar & 3$\%$ & no & \\
25 & $2$~bar & 4$\%$ & no & \\
26 & $2$~bar & 6$\%$ & no & \\
27 & $2$~bar & 8$\%$ & no & \\
28 & $2$~bar & 10$\%$ & no & \\
29 & $2$~bar & 15$\%$ & no & \\
30 & $2$~bar & 4$\%$ & yes (100m GEL) & \\
31 & $2$~bar & 15$\%$ & yes (100m GEL) & \\
\hline
\end{tabular} 
\label{reducing_experiments} 
\end{table*}

\section{Results}
\label{section_results}

\subsection{How much hydrogen is required to warm early Mars?}
\label{reducing_section_amount_hydrogen}

An important aspect of the feasibility of the H$_2$ greenhouse warming scenario is to assess whether or not the amount of H$_2$ required to warm early Mars is compatible with the possible H$_2$ production 
rates \citep{Ramirez:2014,Wordsworth:2017,Kamada:2020,Turbet:2020spectro,Hayworth:2020,Wordsworth:2021}. The amount of hydrogen required to "warm" early Mars has been mainly calculated provided that the global mean surface temperature is above the melting point of water i.e. 273~K. In this section, we revise these numbers based on three distinct constraints (described in the three following subsections) derived from our 3-D global climate simulations: (1) the annual mean globally-averaged surface temperature is above 273~K, or (2) surface liquid water must be present, or (3) the surface temperature of the coldest points must be above 273~K to avoid glacier formation (i.e. to solve the so-called equatorial periglacial paradox).

\subsubsection{Impact of the CO$_2$ and H$_2$ mixing ratios on the global mean surface temperature}
\label{section_tsurf1}

\begin{figure*}
\centering 
\includegraphics[width=\columnwidth]{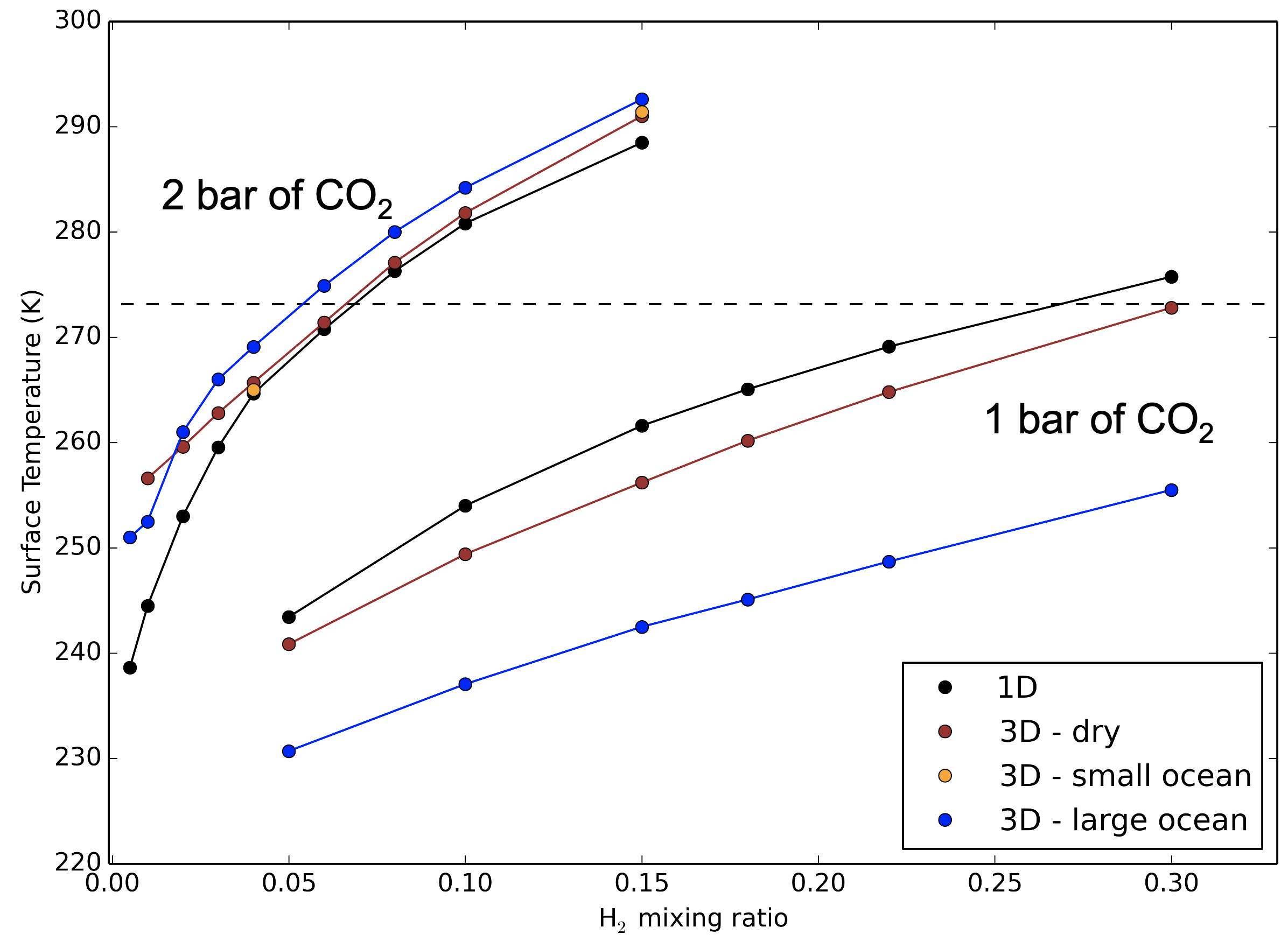}
\caption{Annual global mean surface temperatures (K) as a function of H$_2$ mixing ratio, for two different CO$_2$ partial pressures (1 and 2~bar), and four different types of simulations. These include 1-D simulations (black) based on \citet{Turbet:2020spectro}, 3-D simulations assuming a dry (i.e., low water content) planet (brown), a planet with a small 100~m GEL ocean (orange), and a planet with a large 550~m ocean (blue).}
\label{fig_tsurf_compare}
\end{figure*}

The surface temperature is controlled by a large number of factors including (but not limited to) greenhouse effect, Rayleigh scattering and heat redistribution by CO$_2$, H$_2$ and water ; radiative effects of clouds ; surface albedo changes. Fig~\ref{fig_tsurf_compare} summarizes the evolution of the annual-mean globally averaged surface temperature as a function of H$_2$ mixing ratio, for two different CO$_2$ partial pressures (1 and 2~bar), and three different water contents (dry, 100~m GEL ocean and 550~m GEL ocean). Two families of solutions appear depending on whether the CO$_2$ partial pressure of CO$_2$ is equal to 1 or 2~bar. Simulations at 1~bar require about 4~$\times$ more hydrogen than simulations at 2~bar to reach a surface temperature of 273~K. This stems mostly from the fact that the CO$_2$-H$_2$ collision-induced absorption cross-section is proportional to p$_{\text{CO2}}$~$\times$~p$_{\text{H2}}$~=~p$_{\text{CO2}}^2$~$\times$~x$_{\text{H2}}$, which is also $\sim$~4~$\times$ higher in the 2~bar simulation than in the 1~bar simulation.

\begin{figure*}
\centering 
\subfloat[1~bar of CO$_2$, 550~m GEL ocean]{
\label{subfig:ice_melt_1barCO2_ocean}
\includegraphics[width=\columnwidth]{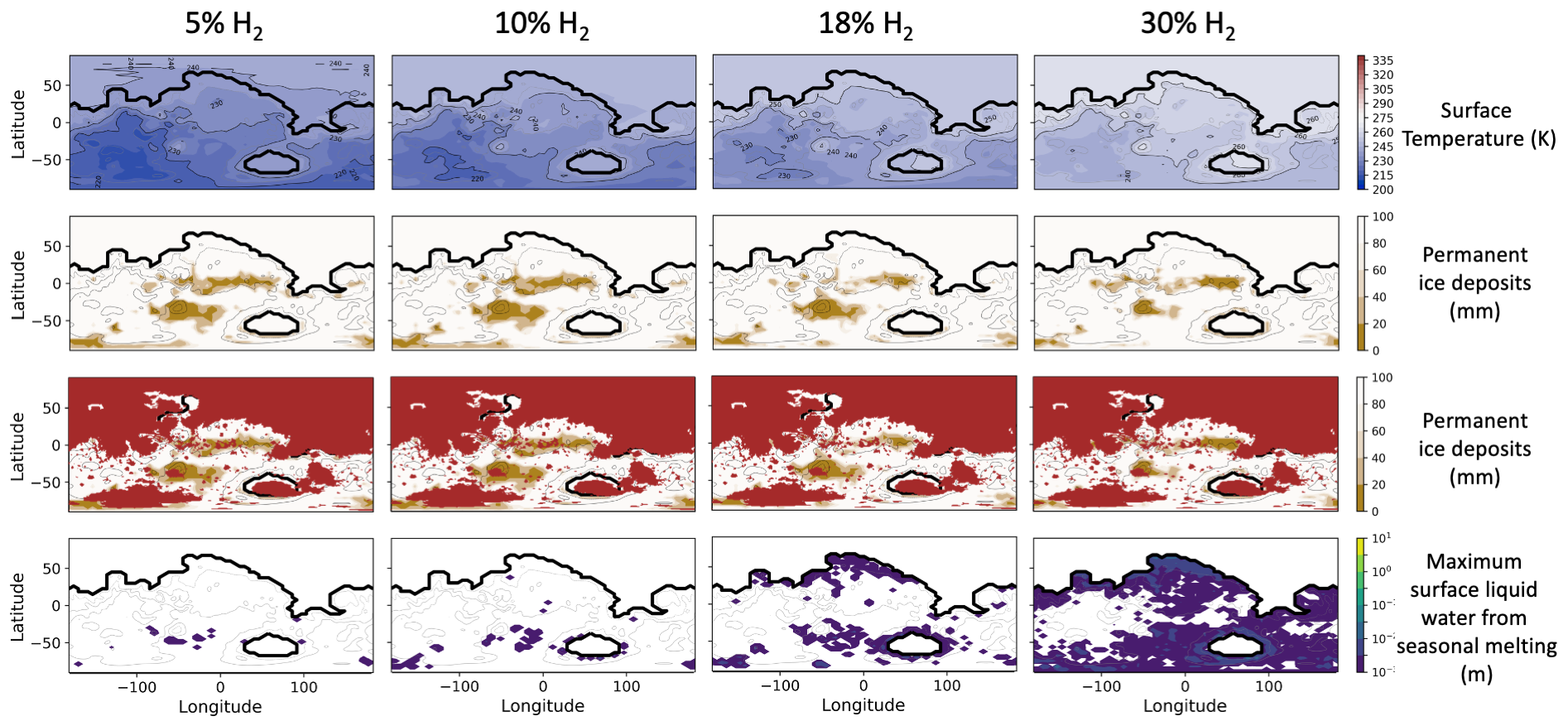}}
\\
\subfloat[1~bar of CO$_2$, no ocean]{
\label{subfig:ice_melt_1barCO2_dry}
\includegraphics[width=\columnwidth]{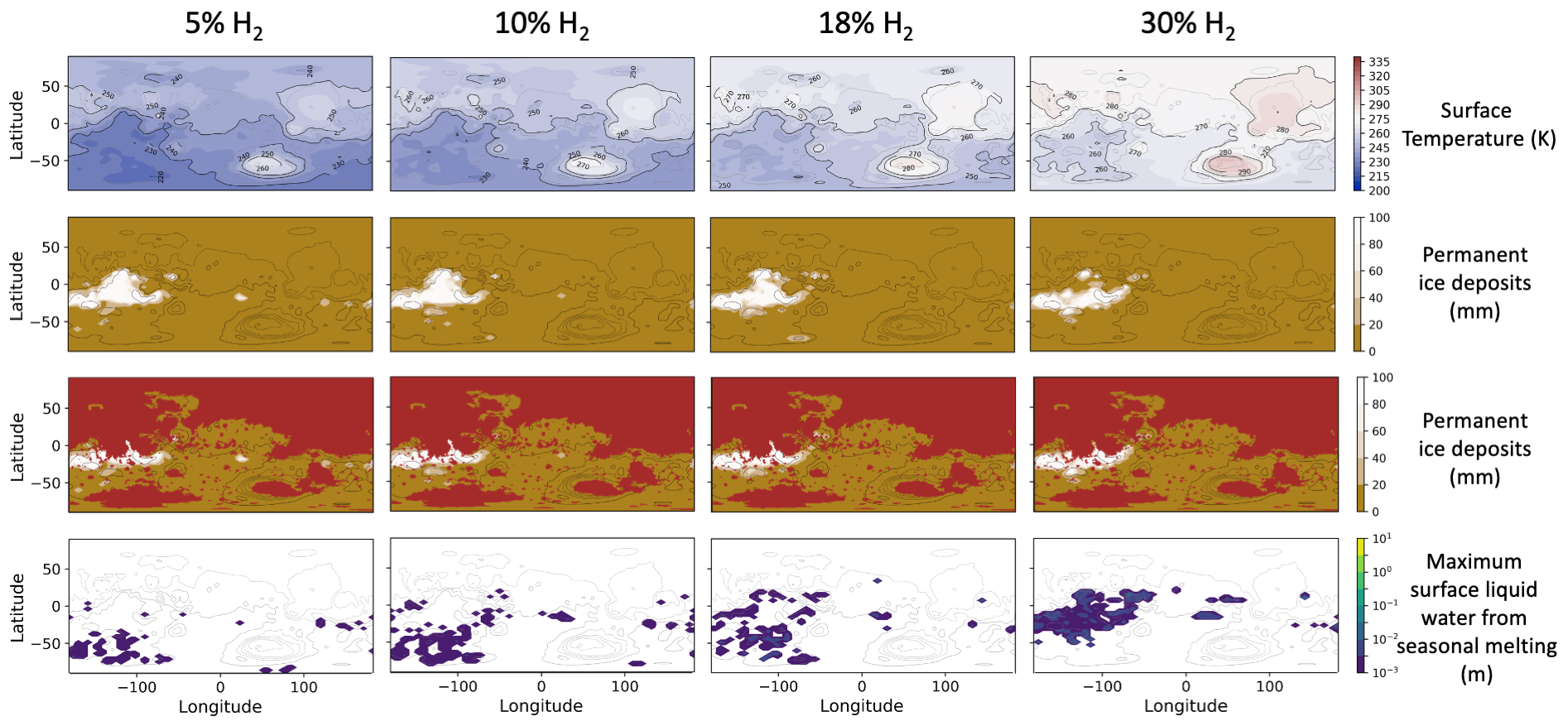}}
\caption{Annual mean surface temperatures (first row), permanent ice deposits (second row), permanent ice deposits filtered from post-Noachian terrains (third row) following \cite{Tanaka:2014}, and maximum surface liquid water produced by seasonal melting (fourth row) maps. Maps were calculated for a 1~bar CO$_2$-dominated atmosphere with various H$_2$ contents (5, 10, 18 and 30$\%$), for a large water reservoir (upper panel) and for a small water reservoir (lower panel). For the simulations without ocean, the maximum amount of water than can be trapped in water ice glaciers is $\sim$~280/210/120/40~m for hydrogen mixing ratio of 5/10/18/30$\%$, respectively. For the simulations with ocean, this amount is $\sim$~1500/1300/1000/600~m for hydrogen mixing ratio of 5/10/18/30$\%$, respectively. These maximum amounts are calculated assuming that glaciers stop thickening when they experience basal melting (``basal melting limit''), using a geothermal heat flux equal to 50~mW~m$^{-2}$.
\label{fig:ice_melt_1barCO2}}
\end{figure*}

\begin{figure*}
\centering 
\subfloat[2~bar of CO$_2$, 550~m GEL ocean]{
\label{subfig:ice_melt_2barCO2_ocean}
\includegraphics[width=\columnwidth]{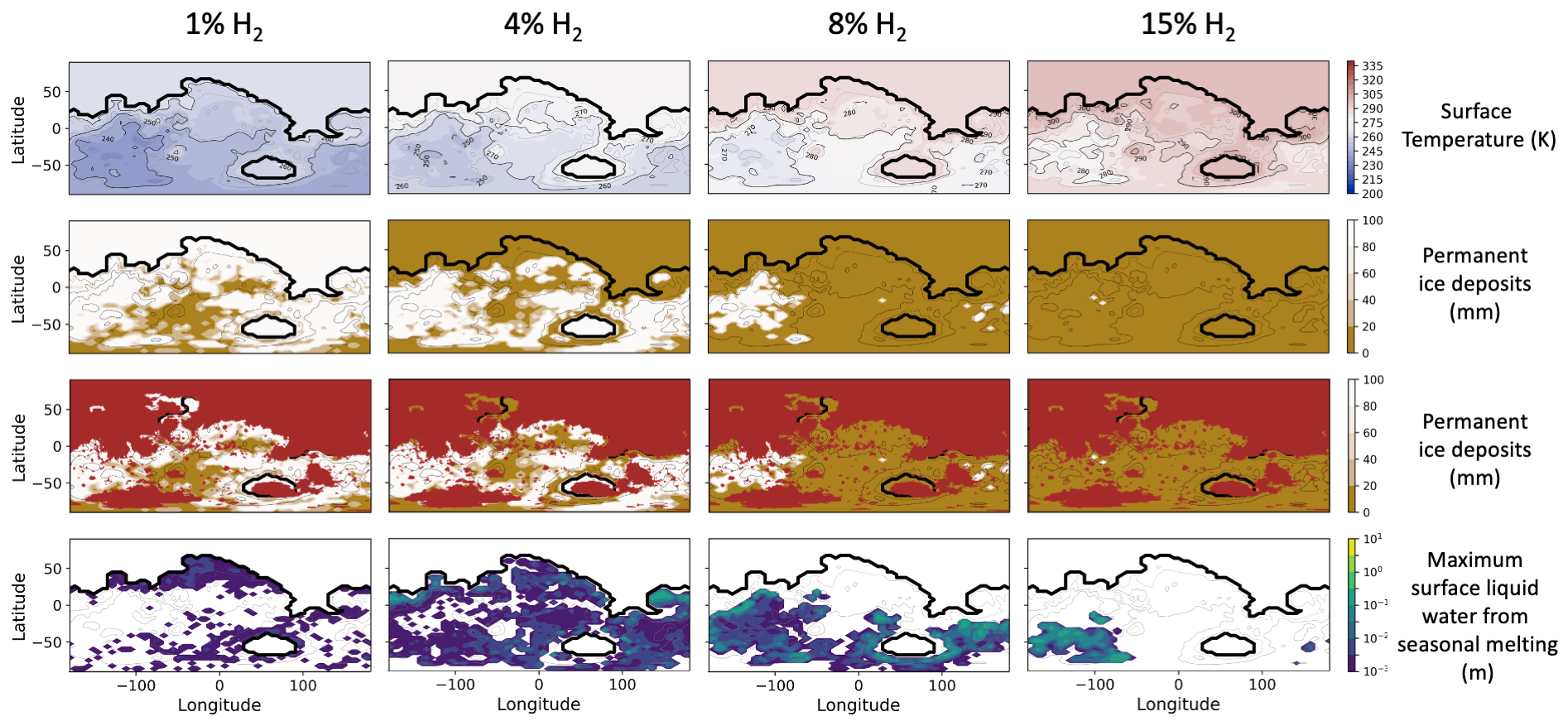}}
\\
\subfloat[2~bar of CO$_2$, no ocean]{
\label{subfig:ice_melt_2barCO2_dry}
\includegraphics[width=\columnwidth]{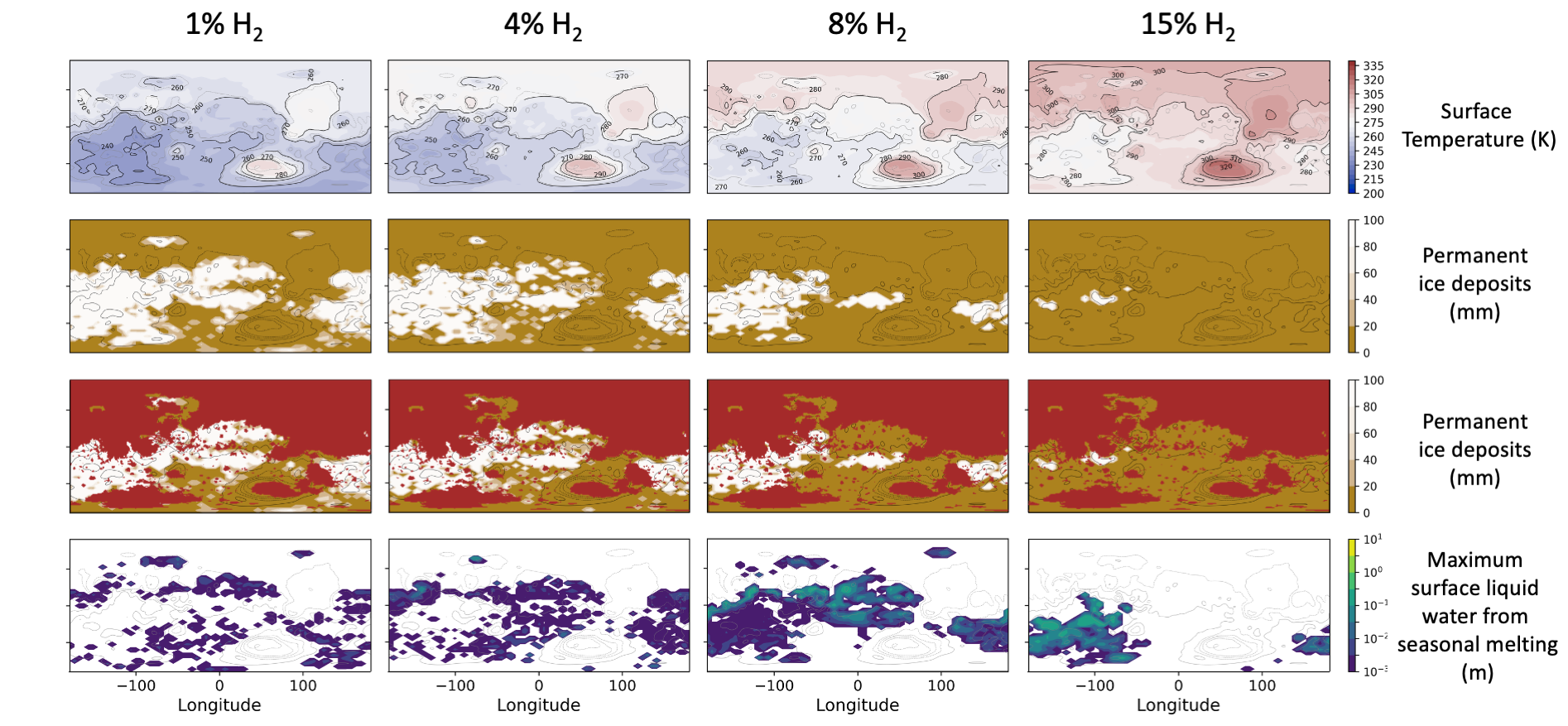}}
\caption{Same as Figure~\ref{fig:ice_melt_1barCO2} but for a 2~bar CO$_2$-dominated atmosphere with H$_2$ contents of 1, 4, 8 and 15$\%$. For the simulations without ocean,  the maximum amount of water than can be trapped in water ice glaciers (basal melting limit) is $\sim$~480/220/60/0~m for hydrogen mixing ratio of 5/10/18/30$\%$, respectively. For the simulations with ocean, this amount is $\sim$~630/200/20/0~m for hydrogen mixing ratio of 5/10/18/30$\%$, respectively.
\label{fig:ice_melt_2barCO2}}
\end{figure*}

We first focus on the 1~bar family solution, which is the most plausible solution given the constraints on the past atmospheric pressure of Mars derived from crater counting \citep{Kite:2014} and estimates of atmospheric escape rates \citep{Lillis:2017,Dong:2018mars}. For the 1~bar CO$_2$-dominated atmosphere, globally-averaged surface temperatures in 3-D simulations are generally colder than calculated with a 1-D model \citep{Turbet:2020spectro} with the same atmospheric composition. This is mainly due to the increase of the planetary albedo linked to the reflectivity of ice deposits accumulating on the surface, as illustrated in Fig~\ref{subfig:ice_melt_1barCO2_ocean}, second row (and to a lesser extent in Fig~\ref{subfig:ice_melt_1barCO2_dry}, second row). This effect is not included in the 1-D model that assumes a fixed surface albedo equal to 0.2 \citep{Turbet:2020spectro}. The difference on the surface temperature is strongest for the ocean simulations (lower blue line in Fig~\ref{fig_tsurf_compare}). Our 3-D simulations indicate in fact that water that evaporates from the ocean is initially deposited almost everywhere on the planet. As the planet cools down, water freezes on the elevated terrains, which increases the albedo, which decreases the mean temperature, and induces even more freezing (e.g. the freezing of the ocean). This is the ice albedo positive feedback. As illustrated by Fig~\ref{subfig:ice_melt_1barCO2_ocean} (second row), the planet is rapidly covered by ice (i.e. becomes a "Snowball Mars") and the mean surface temperature drops to really low values. Note that this effect could possibly be taken into account in 1-D climate models by adjusting/parameterizing the surface albedo \citep{Ramirez:2018craddock}, although the presence of clouds make the estimation complex. 
For scenarios with a small water content (lower red line in Fig~\ref{fig_tsurf_compare}), the difference in surface temperature between the 1D and 3D models is much less pronounced. This is due to the fact that the ice albedo feedback is much less pronounced due to limited ice surface coverage, as illustrated by Fig~\ref{subfig:ice_melt_1barCO2_dry} (second row). Ice deposits accumulate in fact on the most elevated terrains of Mars, the cold traps, due to the adiabatic cooling mechanism \citep{Forget:2013}, in agreement with \citet{Wordsworth:2013}.

\begin{figure*}
\centering 
\includegraphics[width=0.9\columnwidth]{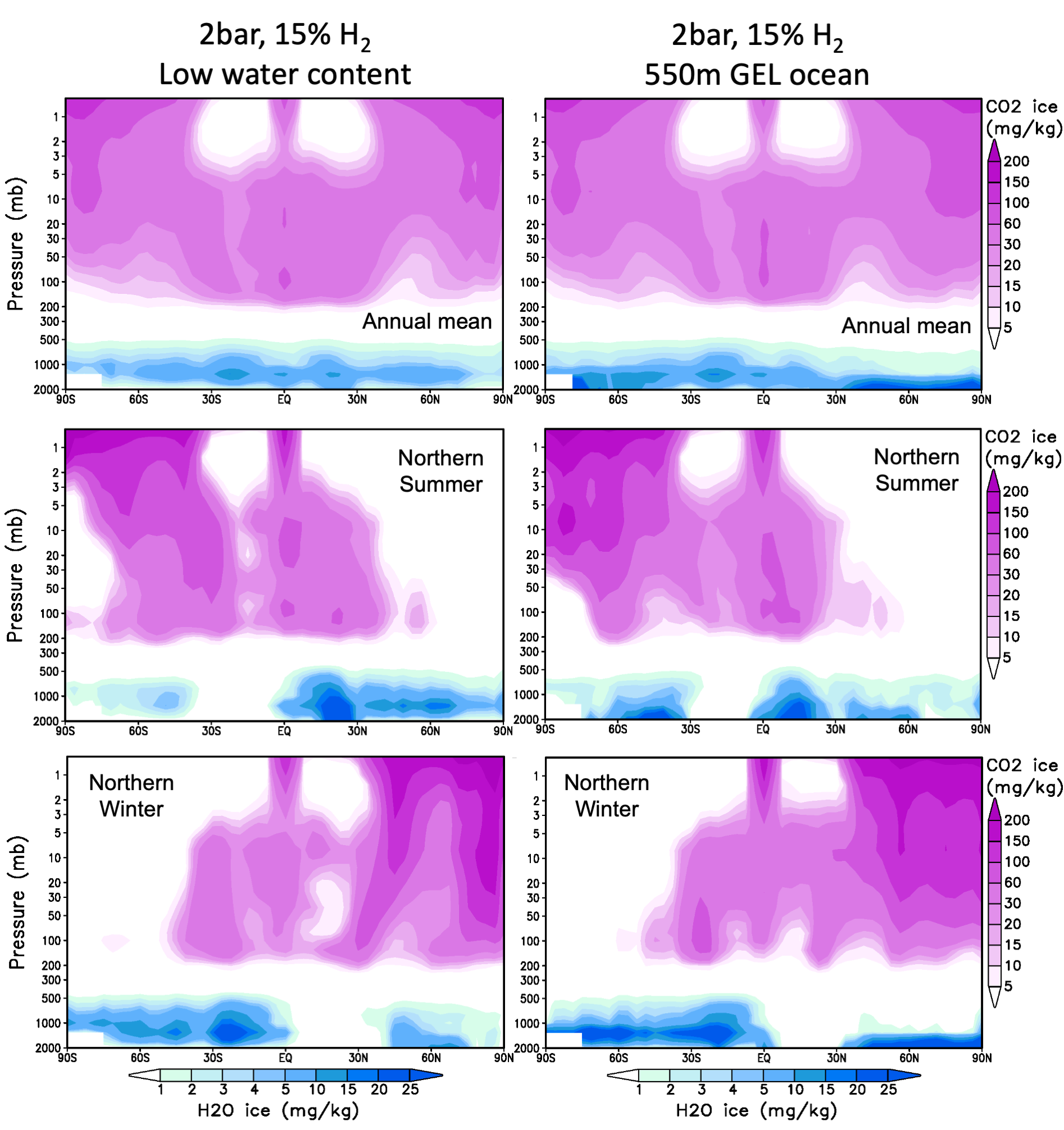}
\caption{Meridional means of the H$_2$O and CO$_2$ ice cloud distributions, for a low water content (water content below a few meters GEL ; left column) and a large northern ocean (about 500 meters GEL ; right column), for 1 Martian year average (top row), 10 Martian days average near the Northern Summer solstice (middle row) and 10 Martian days average near the Northern Winter solstice (bottom row). The figure is computed for atmospheres made of 2~bar of CO$_2$ and 15$\%$ of H$_2$.}
\label{fig_cloud_distributions}
\end{figure*}

We now focus on the 2~bar family solution, which has the advantage (over the 1~bar simulations) of requiring much less H$_2$ to yield temperate temperatures. For the 2~bar CO$_2$-dominated atmosphere, globally-averaged surface temperatures in 3-D simulations are in fact generally warmer than calculated with a 1-D model \citep{Turbet:2020spectro} with the same atmospheric composition. This occurs in spite of the fact that the coverage of surface ice is found to be relatively large (see Figs.~\ref{subfig:ice_melt_2barCO2_ocean} and ~\ref{subfig:ice_melt_2barCO2_dry}, second row) at least for low H$_2$ mixing ratios.
In fact, we identify that the most significant difference (with respect to the 1~bar simulations) is that the greenhouse effect of CO$_2$ ice clouds (not included in 1-D calculations) is much more pronounced here (see Fig~\ref{fig_cloud_distributions}). This is in agreement with the results of \citealt{Forget:2013} (see Figs.~1 and 2) showing that the greenhouse effect of CO$_2$ ice clouds significantly increases with CO$_2$ partial pressure, with an optimum warming near 2~bar. Fig.~\ref{fig_cloud_distributions} illustrates that CO$_2$ ice clouds form seasonally -- in Winter -- in the upper part of the atmosphere (at pressures typically lower than 0.2~bar) where they backscatter efficiently the thermal emission of the surface and the lowermost part of the atmosphere, while having a limited impact on the planetary albedo.

The H$_2$ mixing ratios needed to raise the annual-mean globally averaged surface temperature above 273~K are summarized in Table~\ref{table_cia_tsurf_MAT_273K}. The H$_2$ mixing ratios are all significantly higher than those calculated in \citet{Kamada:2020}, mostly because they used the older CIA coefficients from \citet{Wordsworth:2017}. Moreover, we note some qualitative discrepancies with the simulations presented in \citet{Kamada:2020}. In both the 1~bar and 2~bar families of solutions presented in our work, we observe that the annual-mean globally averaged surface temperature is lower in the ocean case than in the dry case when the planet is cold (cold enough to freeze most surface water), and the result is the opposite when the planet is warm (warm enough so that surface ice deposits are limited in size), due to the ice albedo feedback. This behaviour is not found by \citet{Kamada:2020}, where annual-mean globally averaged surface temperature for their "Aqua-Mars" simulations (i.e. with oceans) is always higher than that of their "Dry-Mars" or "Moist-Mars" (i.e. without oceans), whatever the CO$_2$ and H$_2$ partial pressures considered.

\begin{table*}
\centering
\begin{tabular}{lcl}
   Amount of CO$_2$ & 1~bar & 2~bar \\
   \hline
   1D model \citep{Turbet:2020spectro} & 27$\%$ & 7$\%$ \\
   3D model with large oceans & $>$30$\%$ & 5.5$\%$ \\
   3D model with no oceans & 30$\%$ & 6.5$\%$ \\
   \hline
\end{tabular}
\caption{H$_2$ volume mixing ratio (in $\%$) required to warm the annual-mean globally averaged surface temperature of early Mars above 273K, depending on the assumed CO$_2$ partial pressure (1 or 2~bar), calculated with the LMD Early Mars Generic GCM including the CO$_2$-H$_2$ collision-induced absorption of \citet{Turbet:2020spectro}.}
\label{table_cia_tsurf_MAT_273K}
\end{table*}

\subsubsection{Surface temperature above 273~K is not always a sufficient condition for the presence of surface liquid water}

The fact that the global mean surface temperature of a planet is above the melting point of water (273~K) does not necessarily mean that liquid water exists on its surface. On planets with a limited water reservoir, water is indeed expected to migrate to the coldest points of the planets \citep{Abe:2011,Menou:2013,Leconte:2013aa,Turbet:2016}, i.e. to the polar regions on present-day Mars and Earth, and possibly to the southern highlands on early Mars endowed with a thick CO$_2$ atmosphere \citep{Wordsworth:2013,Turbet:2017icarus}, due to the adiabatic cooling mechanism. Preferential water accumulation in the coldest points of the planet (southern highlands) is also clearly visible in Figs.~\ref{subfig:ice_melt_1barCO2_dry} and ~\ref{subfig:ice_melt_2barCO2_dry}, second row.

\begin{table*}
\centering
\begin{tabular}{lccl}
   & Amount of CO$_2$ & 1~bar & 2~bar \\
   \hline
   condition 1 & 3D model with large oceans & 18$\%$ & 1$\%$ \\
   (melt production $>$1cm) & 3D model with no oceans & 5$\%$ & 1$\%$ \\
   \hline
   condition 2 & 3D model with large oceans & $>$30$\%$ & 3$\%$ \\
   (perennial surface liquid water on land) & 3D model with no oceans & $>$30$\%$ & 6$\%$ \\
   \hline
   condition 3 & 3D model with large oceans & $>$30$\%$ & 3$\%$ \\
   (ocean/lakes not fully ice-covered) &  & & \\
   \hline
\end{tabular}
\caption{H$_2$ volume mixing ratio (in $\%$) required to have perennial surface liquid water, depending on the assumed CO$_2$ partial pressure (1 or 2~bar), and for three distinct conditions (see main text).}
\label{table_cia_tsurf_liquid_water}
\end{table*}

We evaluated in our simulations three distinct conditions (in H$_2$ mixing ratio) for the presence of surface liquid water: (condition 1) the total annual production of surface liquid water (most likely by melting) is larger than 1~cm in at least one GCM grid ; (condition 2) liquid water is continuously present on the surface in at least one continental GCM grid ; (condition 3) oceans/lakes are not entirely covered by ice. The results are summarized in Table~\ref{table_cia_tsurf_liquid_water}. The first condition correspond to a very small amount of water in comparison to the estimated amount needed to carve the valley networks \citep{Luo:2017,Rosenberg:2019,Luo:2020}. It is easier to achieve (i.e., require less H$_2$) than the third, which itself is easier to achieve than the second. In most simulations, these three conditions are also easier to achieve than having the global mean surface temperature above 273~K. 

Simulations with 1 bar of CO$_2$ (which is also the most likely scenario family as hypothesized in \citealt{Wordsworth:2013} and \citealt{Wordsworth:2016}) require extremely high hydrogen mixing ratios ($\ge$30$\%$) not only to raise the global mean surface temperature above 273~K (see Table~\ref{table_cia_tsurf_MAT_273K}), but also to allow the presence of perennial surface liquid water on land (see Table~\ref{table_cia_tsurf_liquid_water}). As detailed in Section~\ref{section_tsurf1}, this stems from the fact that compared with 1-D numerical climate calculations with up-to-date CO$_2$+H$_2$ CIA \citep{Turbet:2020spectro}, (1) in 3-D simulations with an ocean, a strong ice albedo feedback takes place, which produces a strong cooling ; and (2) in 3-D simulations with a low water content, water tends to accumulate on the elevated southern highland terrains, which are also the coldest terrains of the planet, and is thus more difficult to melt.


\subsubsection{Constraints from the equatorial periglacial paradox}

It has been argued that the presence (in 3D GCM simulations) of water ice glaciers in the Martian highlands is an issue, because there is no clear evidence of widespread glacial erosion on these terrains \citep{Ramirez:2018craddock}. This issue is known as the "equatorial periglacial paradox" \citep{Wordsworth:2016}. \citet{Ramirez:2018craddock} advocated that that this so-called "equatorial periglacial paradox" is a piece of evidence against the "icy highland" scenario \citep{Wordsworth:2013,Bouley:2016}, i.e. the possibility that late Noachian valley networks may have formed mainly from the seasonal melting of glacial deposits that accumulated on the southern highlands.

Most of the simulations discussed above predict the formation of permanent water ice reservoirs. Even simulations with global mean surface temperature well above $0^\circ$C also show water ice accumulation, at least in the southern highlands. In such cases, glacial activity and thus erosion are likely to be amplified due to higher surface temperatures (at least, compared with colder simulations, with lower H$_2$ mixing ratios), as illustrated in Figs.~\ref{subfig:ice_melt_1barCO2_ocean},~\ref{subfig:ice_melt_1barCO2_dry},~\ref{subfig:ice_melt_2barCO2_ocean} and ~\ref{subfig:ice_melt_2barCO2_dry}, first row.

We evaluated in our simulations two distinct conditions (in H$_2$ mixing ratio) that may allow to overcome the equatorial periglacial paradox: (condition 1) no ice accumulation occurs in the GCM simulation (<1~m GEL) ; (condition 2) no ice accumulation occurs in the GCM grids corresponding to Noachian (or older) terrains (<1~m GEL). The results are summarized in Table~\ref{table_cia_tsurf_glacier}. We found that the H$_2$ mixing ratios required to reach one or the other of the condition is the same. This stems from the fact that the adiabatic cooling mechanism tends to form ice deposits on the southern highlands \citep{Wordsworth:2013,Bouley:2016,Turbet:2017icarus} which are mostly Noachian terrains. In all configurations (i.e. for all water contents and CO$_2$ pressures considered), the absence of ice deposits on Noachian terrains requires significantly more hydrogen (e.g., more than twice for the simulation with 2~bar of CO$_2$ and large oceans) than having the global mean surface temperature above 273~K. This is due to the fact that preventing glacier formation requires having the coldest points of the planets above 273~K, which is a much more stringent condition.

In any case, even if the climate was warm enough to avoid the formation of glaciers for some time, it seems difficult to avoid glacier formation when the climate transitioned to colder epochs (for example when the H$_2$ mixing ratio was much lower), as discussed in \citet{Wordsworth:2016}.

One possibility to resolve this apparent paradox is that the glaciers could have had a cold base, which would drastically reduce glacial erosion on the highlands \citep{Fastook:2015}. Whether glaciers had a cold base or not depends on the water content available at the surface and in the atmosphere at the Late Noachian epoch, as well as the surface temperature in the highlands and the geothermal heat flux. In warm simulations (i.e. with large amount of H$_2$) the area over which glaciers form gets smaller (with respect to the cold simulations with low amount of H$_2$). However it is also expected that -- for a fixed amount of surface water -- glaciers should have a warmer base, and thus be more likely to achieve glacial erosion. Counterintuitively, and unless the atmosphere is warm enough to totally prevent the formation of glaciers, warm climates should lead to much more extensive glacial erosion than cold climates.

Another possibility to solve the paradox is that traces of glacier flows are present but so altered that they have not yet been detected. Recent works \citep{Bouquety:2019,Bouquety:2020} reported morphometric evidence of glaciation (glacial valleys and cirques) in Terra Sabaea (a southern highland terrain), dating back from late Noachian/early Hesperian. Searches for traces of glacier erosion should now be carried out on a wide range of terrains dating from the late Noachian/early Hesperian period, especially on the most elevated ones such as Terra Sirenum, where most of our GCM simulations predict preferential ice accumulation.

\begin{table*}
\centering
\begin{tabular}{lccl}
    & Amount of CO$_2$ & 1~bar & 2~bar \\
   \hline
   condition 1 & 3D model with large oceans & $>$30$\%$ & 10$\%$ \\
   (no ice) & 3D model with no oceans & $>$30$\%$ & 10$\%$ \\
   \hline
   condition 2 & 3D model with large oceans & $>$30$\%$ & 10$\%$ \\
   (no ice on Noachian terrains) & 3D model with no oceans & $>$30$\%$ & 10$\%$ \\
   \hline
\end{tabular}
\caption{H$_2$ volume mixing ratio (in $\%$) required to possibly overcome the equatorial periglacial paradox, depending on the assumed CO$_2$ partial pressure (1 or 2~bar), and for two distinct conditions (see main text).}
\label{table_cia_tsurf_glacier}
\end{table*}


\begin{figure*}
\centering 
\includegraphics[width=\columnwidth]{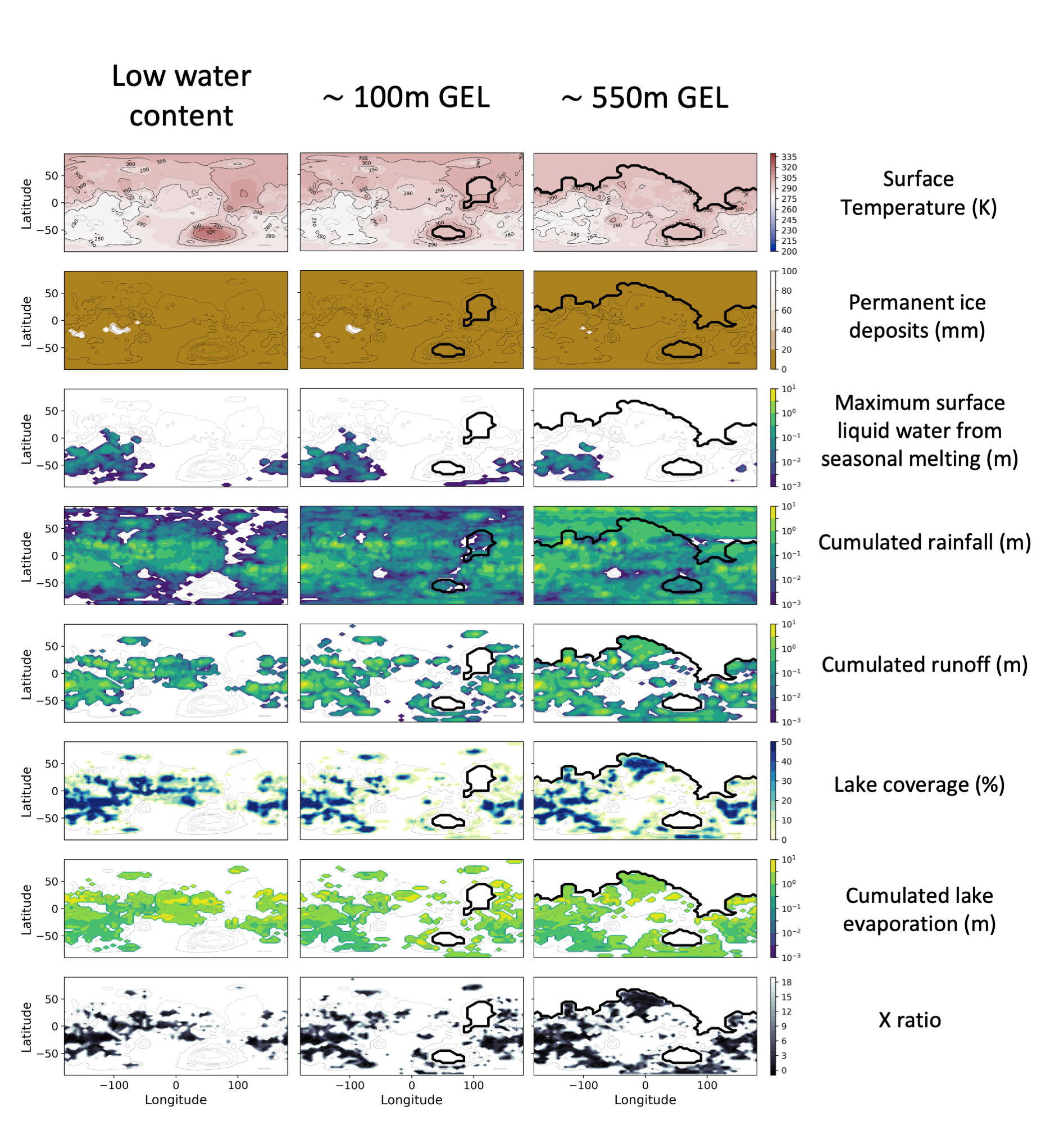}
\caption{Annual-mean average quantities (surface temperatures, permanent ice deposits, maximum surface liquid water produced by seasonal melting, cumulated rainfall, cumulated runoff, lake coverage, cumulated lake evaporation and aridity index ``X-ratio'') for atmospheres made of 2~bar of CO$_2$ and 15$\%$ of H$_2$, for various total water inventories. Note that the cumulated lake evaporation must be multiplied by the lake coverage $\alpha_{\text{lake}}$ to obtain the cumulated evaporation for the GCM grid. The ``Low water content simulations correspond to a range of water content of a few meters GEL, depending on how thick the glaciers are (the exact value has no impact on the modeled climate; see Figures ~\ref{fig:ice_melt_1barCO2} and~\ref{fig:ice_melt_2barCO2}).
See Fig~\ref{fig_full_2bar_15pH2_MASKED} for a comparison with the positions of Noachian terrains \citep{Tanaka:2014} and known valley networks \citep{Bouley:2016}.}
\label{fig_full_2bar_15pH2}
\end{figure*}

\begin{figure*}
\centering 
\includegraphics[width=\columnwidth]{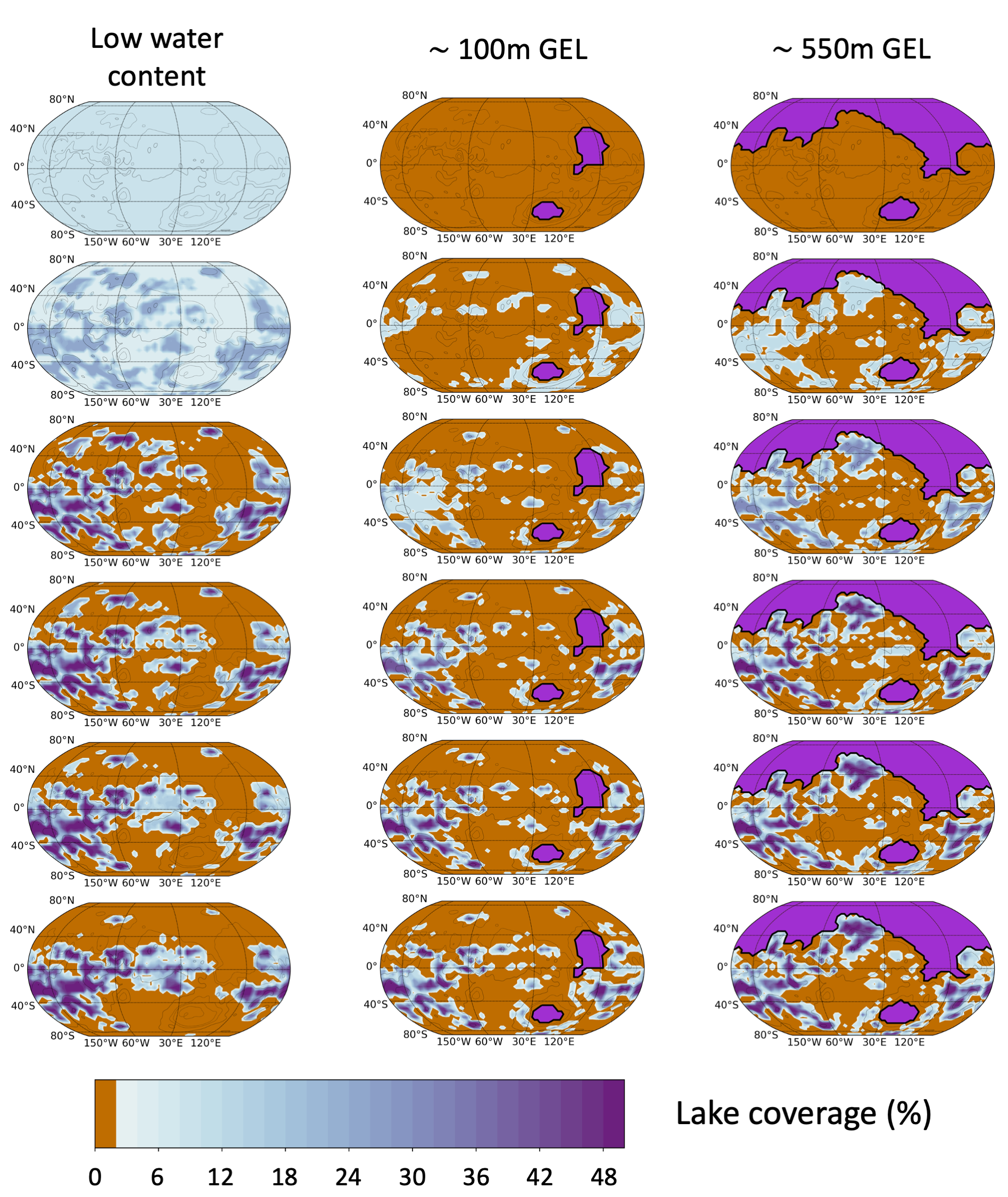}
\caption{Evolution of lake coverage (in $\%$) at timesteps = 0, 1, 4, 9, 15 and 30 for the 2~bar-CO$_2$, 15$\%$-H$_2$ simulation presented in Figure~\ref{fig_full_2bar_15pH2}.
\label{fig_lake_evol}}
\end{figure*}

\subsection{The nature of the hydrologic cycle in warm simulations}
\label{reducing_section_warm_climates}

The total surface/near-surface water content available during the Late Noachian is still debated \citep{Wordsworth:2016}. 
\citet{Carr:2015} estimated that the total reservoir could have been as low as $\sim$~24~m GEL, based on water loss/gain budget during the subsequent Hesperian and Amazonian epochs. \citet{Villanueva:2015} calculated that the late Noachian water content was $\sim$~60~m GEL, based on present-day Mars isotopic D/H ratio. Eventually, \citet{Diachille:2010} argued that the total water inventory was $\sim$~550m GEL, based on possible ancient ocean shorelines determined by the position of deltas.

We explore in this section how the nature of the early Mars hydrologic cycle under a strong greenhouse warming (strong enough for rain to be produced and lakes to be formed) changes depending on the assumed total water inventory. For this we designed a series of 3-D numerical climate simulations (see Table~\ref{reducing_experiments}) to explore possibles climates ranging from warm $\&$ arid to warm $\&$ wet conditions. We focus specifically on simulations with 2~bar CO$_2$-dominated atmospheres as they allow warm climates to be achieved with a relatively reasonable H$_2$ mixing ratio (see Tables~\ref{table_cia_tsurf_MAT_273K}, \ref{table_cia_tsurf_liquid_water} and \ref{table_cia_tsurf_glacier}).

\subsubsection{Warm and arid scenarios}

Warm $\&$ arid scenarios are cases where the total water content on Mars was low enough so that no oceans formed, as depicted in Fig~\ref{fig_full_2bar_15pH2} (column~1).

\citet{Wordsworth:2013} demonstrated using 3-D Global climate simulations that - in cold and arid scenarios - water should migrate to the cold traps of the planet (e.g. to the southern highlands) and accumulate there as ice, through the mechanism of adiabatic cooling and subsequent condensation. We find the same behaviour in our simulations with low water inventory and moderate greenhouse warming (see e.g. Figs~\ref{fig:ice_melt_1barCO2}b and \ref{fig:ice_melt_2barCO2}b). However, in our simulations, the size of the permanent ice reservoirs is more limited, for two reasons:
\begin{itemize}
\item The number of locations on Mars where the annual mean surface temperature is below 0$^\circ$C is reduced. Therefore, the number of possible locations where water can accumulate as ice is also reduced.
\item The maximum thickness of the permanent ice reservoirs is also reduced, because of the basal melting limit (defined by the maximum thickness of ice deposits before they start to melt at their base) is more restrictive at higher surface temperatures.
\end{itemize}

In contrast with the predictions of \citet{Wordsworth:2016}, we actually find a similar behavior in simulations with a strong greenhouse warming (shown in Fig~\ref{fig_full_2bar_15pH2}, column~1), i.e. strong enough to prevent the formation or permanent ice deposits. In these warm $\&$   arid simulations, water was initially uniformly distributed in lakes all over the planet (with an arbitrary mean lake surface coverage of 10$\%$ ; this corresponds to a total water content of 4~m GEL). As illustrated in Fig~\ref{fig_lake_evol} (column~1), water progressively migrates to regions where the {precipitation~-~evaporation} budget is favorable, i.e. the cold traps. Because of the adiabatic cooling process, the cold traps in our warm simulations are also located in the southern highlands, as calculated by \citet{Wordsworth:2013} in much colder configurations. As a result, water rapidly migrates to the southern highlands where it accumulates - in liquid form - in impact crater lakes (see Fig~\ref{fig_full_2bar_15pH2}, column~1, row~6 ; or Fig~\ref{fig_lake_evol}, column~1, row~6).

Once water is stabilized in southern highlands impact crater lakes, precipitation (rainfall) occurs close to the reservoirs of water (lakes), i.e. also in the southern highlands (see Fig~\ref{fig_full_2bar_15pH2}, column~1, row~4). A good predictor of the regions of valley networks formation is the annual mean cumulated runoff (see Fig~\ref{fig_full_2bar_15pH2}, column~1, row~5) that shows that valley networks - in this scenario - would also form in the southern highlands (rainfall are less indicative because in our model it often rain without yielding runoff because the soil is not saturated; see section~\ref{sc:continental}). Both lake formation and runoff (initiated by rainfall) are widespread and mostly located in the southern highlands, were both impact crater lakes and valley networks have been observed \citep{Fassett:2008b, Hynek:2010}. We note a relatively strong runoff in Arabia Terra, where no valley networks have been directly observed. However, \cite{Davis:2016} identified extensive networks of sinuous ridges, which they interpreted as inverted Noachian fluvial channels that would have been buried and then exhumed. Our 3-D simulations are in agreement with this interpretation. Detailed comparisons with regions of valley network formations \citep{Bouley:2016} filtered from post-Noachian terrains \citep{Tanaka:2014} are provided in Fig~\ref{fig_full_2bar_15pH2_MASKED}.

The results of these warm and arid simulations is promising because they show that the adiabatic cooling mechanism can also work in warm climates, leading to extensive valley network formation on the Martian highlands as observed. However, there are two issues in this scenario. First, the simulations explored here correspond to a very low surface water total inventory ($\sim$~4~m GEL) compared to the range of available estimates \citep{Diachille:2010,Villanueva:2015,Carr:2015}. Secondly, a large fraction of the lakes that form in these simulations are overflowing. Most of the lake forming regions are associated to very low aridity X-ratios (see Fig~\ref{fig_full_2bar_15pH2}, column 1, row 8). For a X-ratio lower than $\sim$~2, lakes are expected to overflow \citep{Howard:2007,Matsubara:2011}. Some lakes that reached their maximal size can even have a positive {precipitation~-~evaporation} positive budget. Therefore, in our simulations, a large number of lakes should be open-basin lakes, which seems in contradiction with the observational evidence that closed-basin lakes greatly outnumber open-basin lakes \citep{Wordsworth:2016}. 

\begin{figure*}
\centering 
\includegraphics[width=\columnwidth]{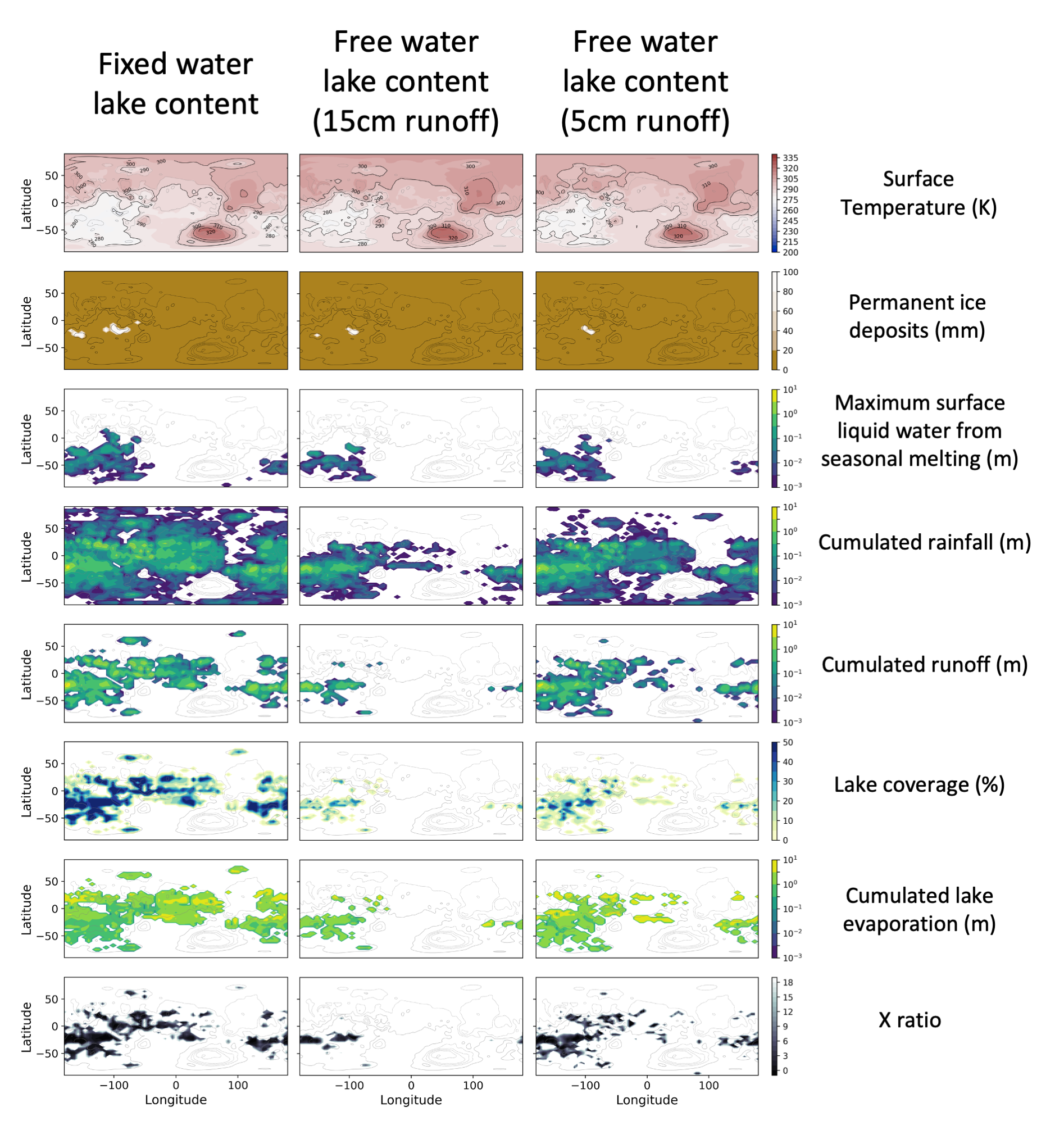}
\caption{Annual-mean average quantities (surface temperatures, permanent ice deposits, maximum surface liquid water produced by seasonal melting, cumulated rainfall, cumulated runoff, lake coverage, cumulated lake evaporation and aridity index ``X-ratio'') for atmospheres made of 2~bar of CO$_2$ and 15$\%$ of H$_2$, for very low surface water inventories. Simulation (shown in left column) was calculated with a water reservoir in the lakes fixed at 4m GEL. Simulations (shown in middle and right columns) were calculated with a water reservoir in the lakes left free, and for two different assumptions: a runoff initiated from 15cm (middle column) or 5cm (right column) surface liquid water.}
\label{fig_full_2bar_15pH2_sensitivity_study}
\end{figure*}

\begin{figure*}
\centering 
\includegraphics[width=\columnwidth]{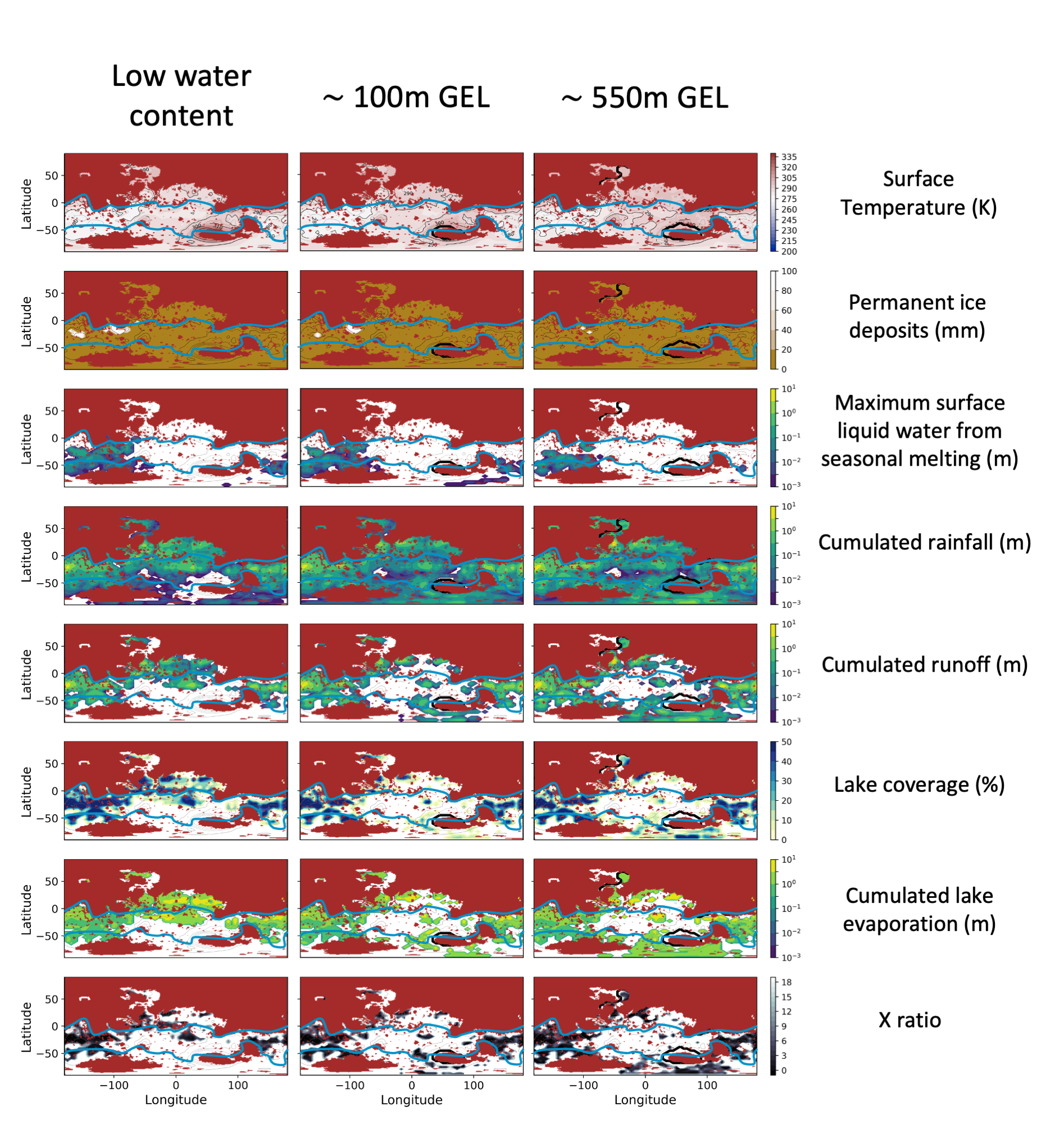}
\caption{Annual-mean average quantities (surface temperatures, permanent ice deposits, maximum surface liquid water produced by seasonal melting, cumulated rainfall, cumulated runoff, lake coverage, cumulated lake evaporation and X-ratio) for atmospheres made of 2~bar of CO$_2$ and 15$\%$ of H$_2$, for various total water inventories. Note that the cumulated lake evaporation must be multiplied by the lake coverage $\alpha_{\text{lake}}$ to obtain the cumulated evaporation for the GCM grid. Here we have filtered (in red) from the maps the terrains younger than the Noachian, based on \cite{Tanaka:2014} geological map in a pre-TPW geometry \cite{Bouley:2016}. We have also added the valley networks known positions from \cite{Bouley:2016}.}
\label{fig_full_2bar_15pH2_MASKED}
\end{figure*}

This second problem might just be a consequence of the simplicity of our hydrological parametrization. Firstly, it does not take into account the large scale transport of water by runoff. One could imagine that a few open-basin lakes feeding rivers able to transport water across several grid meshes could significantly drain and dry a region, preventing there the formation of overflowing lakes. Secondly, in our simulations we forced the hydrological scheme to keep the total amount of liquid water trapped in lakes to a fixed value (4~m GEL). We performed several sensitivity studies (see Fig~\ref{fig_full_2bar_15pH2_sensitivity_study}, columns 2 and 3) where we let the martian lakes freely evolve for a total of 20 timesteps, starting from the initial state presented in Fig~\ref{fig_full_2bar_15pH2} (left column). We carried out several sensitivity tests varying the quantity of liquid water before triggering the runoff (from 1.5~cm to 15~cm) because we had identified this parameter as being the most important in the lake distribution. The results -- presented in Fig~\ref{fig_full_2bar_15pH2_sensitivity_study}, columns 2 and 3 -- show that while the geographical distribution of valley networks remains mostly located on the highlands, the fraction of overflowing lakes is now very small. However, these results were obtained at the expense of not conserving water in lakes.

\subsubsection{Warm and wet scenarios}

Warm $\&$ wet scenarios are cases where the total water content on Mars was high enough for permanent oceans to form, as depicted in Fig~\ref{fig_full_2bar_15pH2} (columns 2 and 3). In the warm and wet simulations, the total amount of water possibly trapped in impact crater lakes is let free to evolve. Water evaporates from the oceans, precipitates on "continental" regions where it can accumulate and form impact crater lakes. 

In these scenarios, a fraction of the available water migrates to the cold traps of the planet, where it accumulates because the evaporation/condensation budget is favorable there. For this reason, our simulations also indicate that impact crater lakes (and thus rainfall) are widespread in the southern highlands (see Fig~\ref{fig_full_2bar_15pH2}, columns 2 and 3, row 6). 

This result actually differs from the prediction of \citet{Wordsworth:2015}, despite the fact that our simulations with ocean (at 550~m GEL) appear to have a very similar setup. The difference is probably due to the fact that \citet{Wordsworth:2015} use a version of the LMD model that could not take into account the formation of lakes. The lakes maintains a significant reservoir of liquid water throughout the year and increase the rainfall and runoff rate on the highlands. Another key difference driving the precipitation patterns may be topography (see Fig~\ref{figure_reducing_topo_earlymars}). In our simulations we used the pre-True Polar Wander (pre-TPW) topography from \citet{Bouley:2016} whereas \citet{Wordsworth:2015} assumed the present day topography, corrected to account the presence of an ocean. \citet{Wordsworth:2015} reported a lack of precipitation in the Margaritifer Sinus region (east of Tharsis) and demonstrated that this absence of precipitation was due to a rain shadow effect: Westerly winds over Tharsis produce adiabatic cooling and thus precipitation on the western flank of Tharsis, which dries out the air and thus reduces precipitation on the east of Tharsis \citep{Wordsworth:2015}. Because the Tharsis volcanic province has been removed from the topography in our simulations, this effect disappears and we record rainfall in the Margaritifer Sinus region in our simulations (see e.g. Fig~\ref{fig_full_2bar_15pH2}, columns 2 and 3, row 4).

In addition, orographic precipitation (rainfall) is produced near the shorelines of the oceanic reservoirs of water. Subsequently, impact crater lakes are formed in these regions. For instance,  we find that the presence (and coverage) of a liquid water reservoir in Hellas crater dictates whether or not precipitation occurs in the nearby areas. With a low water reservoir in Hellas (or no reservoir at all), precipitation around Hellas and especially in Noachis Terra (where only a few valley networks have been observed) is suppressed. The position of lakes and rainfall events is thus tightly linked to the position of the oceanic reservoirs.

\medskip

More generally, for warm scenarios, precipitation and lake formation patterns can be understood as the sum of two contributions:
\begin{enumerate}
\item A contribution due to the presence of cold traps on the planet. The precipitation/condensation budget will always favor the formation of lakes (and subsequently of precipitation) in the cold traps. This is a very generic process that explains why water accumulates in the cold traps in both cold regimes \citep{Wordsworth:2013} and warm regimes (this work). At high enough obliquity and high enough pressure, because of the adiabatic cooling effect, the cold traps are located in the southern highlands.
\item A contribution due to global reservoirs of water (oceanic regions). Depending on the location and the area of the oceanic reservoirs, the precipitation patterns (and thus the lake formation patterns) are affected. 
\end{enumerate}
This result is promising because it indicates that 1) not only the presence of precipitation and impact crater lakes in the highlands is robust in 3-D global climate simulations, but also 2) the other geographic patterns of precipitation and impact crater lakes formation can possibly be "tuned" in a warm scenario by (i) using an adapted topography (typical of late Noachian Mars ; e.g. removing the Tharsis province), by (ii) adjusting the amount and locations of surface water reservoirs\footnote{The amount and locations of surface water reservoirs are anyway highly uncertain, not only because the total water content is unknown, but also because it depends on subsurface mechanisms of recharge.} (e.g. remove the ocean from Hellas impact crater to avoid valley networks formation in Noachis Terra) and by (iii) adjusting the amount of greenhouse gases (here CO$_2$ and H$_2$).

\section{Conclusions and Discussions}
\label{conclusions_discussions}

In conclusion, arid and semi-arid climates are promising solutions to the early Mars enigma. Using a 3-D numerical climate model adapted to account for the hydrology of impact crater lakes, we demonstrated that precipitation and lake formation patterns deduced from geology can be roughly reproduced by climate models - in warm and arid/semi-arid climates - on three conditions: (1) a strong grenhouse gas (like H$_2$ here) must be added to CO$_2$ in the atmosphere (2) a topography adapted for Late Noachian Mars conditions must be assumed and (3) the amount of available surface water in the system must be adjusted. 

In particular, the presence of valley networks around Margaritifer Sinus can be reconciled with climate models assuming that the main bulk of Tharsis was emplaced after the formation of valley networks, as predicted by \citet{Bouley:2016}. Moreover, the absence of valley networks e.g. around Noachis Terra can be reconciled with climate models assuming that no sea formed in Hellas impact basin. Eventually, the presence of a relatively strong runoff in Arabia Terra is consistent with the detection of networks of inverted fluvial channels \citep{Davis:2016}

However, there are some significant remaining caveats associated with this scenario. Firstly, the total amount of reducing gas (here H$_2$) required to produce the aforementioned fluvial features is likely much higher than predicted in \citet{Ramirez:2017a} because - by increasing order of exigence - (1) of the ice albedo feedback, (2) of the (non-)availability of water in regions of higher temperatures, and (3) of the equatorial periglacial paradox. The levels of H$_2$ required (i) to produce rainfall on the southern highland (Noachian) terrains and (ii) to overcome the equatorial periglacial paradox are several times higher than predicted in \citet{Ramirez:2017a}, using here the revised CO$_2$+H$_2$ CIA estimates of \citet{Turbet:2020spectro}.

Secondly, in our warm simulations (either "dry" or "wet"), precipitation rates tend to be high on the highlands. As a result, most of the impact crater lakes that form on the highlands are overflowing, which seems at odd with the observational evidence that closed-basin lakes outnumbered open-basin lakes \citet{Wordsworth:2016}.

This last caveat could be revisited in more details by improving our model (the hydrological parametrization in particular) and by simulating cases that we could not model in this paper: with obliquities different than 40$^\circ$, with  different CO$_2$ partial pressure and H$_2$ contents, with different CO$_2$ and H$_2$O cloud properties. All these aspects could impact the distribution of precipitation and thus surface runoff.

We encourage future works exploring the impact of warm climates on the hydrologic cycle of Mars to develop a self-coherent treatment of the evolution of water reservoirs. This includes (i) simulating the evolution and migration of large water reservoirs (is an ocean climatically stable in Hellas? in the northern plains? should it be trapped as ice on the highlands? does it depend on initial water reservoir positions?), (ii) simulating the evolution of water ice reservoirs  (are glaciers expected to have a cold or warm base?), (iii) simulating a true temporal evolution of the impact crater lakes to ensure water conservation.

In parallel, we also encourage future works exploring the impact of warm climates to search for optimal sets of parameters (water content and distribution ; CO$_2$ and/or H$_2$ partial pressures ; orbital parameters ; runoff parameterization) to demonstrate whether there is a set of parameters that could optimally fit the Late Noachian valley networks observed distribution.

Combining these two approaches may be key to solving the early Mars enigma.

\section{Acknowledgements}
This project has received funding from the European Union’s Horizon 2020 research and innovation program under the Marie Sklodowska-Curie Grant Agreement No. 832738/ESCAPE. This project has received funding from the European Research Council (ERC) under the European Union’s Horizon 2020 research and innovation program (grant agreement No. 835275/Mars through time). This work has been carried out within the framework of the National Centre of Competence in Research PlanetS supported by the Swiss National Science Foundation. M.T. acknowledges the financial support of the SNSF. M.T. thanks the Gruber Foundation for its generous support to this research. This work was performed using the High-Performance Computing (HPC) resources of Centre Informatique National de l'Enseignement Supérieur (CINES) under the allocations No. A0060110391 and A0080110391 made by Grand Équipement National de Calcul Intensif (GENCI). M.T. thanks Sébastien Lebonnois for granting him a preliminar access to the OCCIGEN supercomputer. A total of $\sim$~400k CPU hours were used for this project on the OCCIGEN supercomputer, resulting in $\sim$~700~kg eq. of CO$_2$ emissions. FF and MT thank the LMD Generic Global Climate team for the teamwork development and improvement of the model.
\bibliography{Thesis}

\begin{thebibliography}{}

\bibitem[{Abe} et~al., 2011]{Abe:2011}
{Abe}, Y., {Abe-Ouchi}, A., {Sleep}, N.~H., and {Zahnle}, K.~J. (2011).
\newblock {Habitable Zone Limits for Dry Planets}.
\newblock {\em Astrobiology}, 11:443--460.

\bibitem[{Baranov} et~al., 2004]{Baranov:2004}
{Baranov}, Y.~I., {Lafferty}, W.~J., and {Fraser}, G.~T. (2004).
\newblock {Infrared spectrum of the continuum and dimer absorption in the
  vicinity of the O $_{2}$ vibrational fundamental in O $_{2}$/CO $_{2}$
  mixtures}.
\newblock {\em Journal of Molecular Spectroscopy}, 228:432--440.

\bibitem[{Bouley} et~al., 2016]{Bouley:2016}
{Bouley}, S., {Baratoux}, D., {Matsuyama}, I., {Forget}, F.,
  {S{\'e}journ{\'e}}, A., {Turbet}, M., and {Costard}, F. (2016).
\newblock {Late Tharsis formation and implications for early Mars}.
\newblock {\em Nature}, 531:344--347.

\bibitem[{Bouquety} et~al., 2020]{Bouquety:2020}
{Bouquety}, A., {Sejourn{\'e}}, A., {Costard}, F., {Bouley}, S., and
  {Leyguarda}, E. (2020).
\newblock {Glacial landscape and paleoglaciation in Terra Sabaea: Evidence for
  a 3.6 Ga polythermal plateau ice cap}.
\newblock {\em Geomorphology}, 350:106858.

\bibitem[{Bouquety} et~al., 2019]{Bouquety:2019}
{Bouquety}, A., {Sejourn{\'e}}, A., {Costard}, F., {Mercier}, D., and {Bouley},
  S. (2019).
\newblock {Morphometric evidence of 3.6 Ga glacial valleys and glacial cirques
  in martian highlands: South of Terra Sabaea}.
\newblock {\em Geomorphology}, 334:91--111.

\bibitem[{Cabrol} and {Grin}, 1999]{Cabrol:1999}
{Cabrol}, N.~A. and {Grin}, E.~A. (1999).
\newblock {Distribution, Classification, and Ages of Martian Impact Crater
  Lakes}.
\newblock {\em Icarus}, 142:160--172.

\bibitem[{Carr}, 1995]{Carr:1995}
{Carr}, M.~H. (1995).
\newblock {The Martain drainage system and the origin of valley networks and
  fretted channels}.
\newblock {\em Journal of Geophysical Research}, 100:7479--7507.

\bibitem[{Carr} and {Head}, 2015]{Carr:2015}
{Carr}, M.~H. and {Head}, J.~W. (2015).
\newblock {Martian surface/near-surface water inventory: Sources, sinks, and
  changes with time}.
\newblock {\em Geophysical Research Letters}, 42:726--732.

\bibitem[{Carter} et~al., 2015]{Carter:2015}
{Carter}, J., {Loizeau}, D., {Mangold}, N., {Poulet}, F., and {Bibring}, J.-P.
  (2015).
\newblock {Widespread surface weathering on early Mars: A case for a warmer and
  wetter climate}.
\newblock {\em Icarus}, 248:373--382.

\bibitem[{Charnay} et~al., 2013]{Charnay:2013}
{Charnay}, B., {Forget}, F., {Wordsworth}, R., {Leconte}, J., {Millour}, E.,
  {Codron}, F., and {Spiga}, A. (2013).
\newblock Exploring the faint young sun problem and the possible climates of
  the archean earth with a 3-d gcm.
\newblock {\em Journal of Geophysical Research : Atmospheres}, 118:414--431.

\bibitem[{Chassefi{\`e}re} et~al., 2016]{Chassefiere:2016}
{Chassefi{\`e}re}, E., {Lasue}, J., {Langlais}, B., and {Quesnel}, Y. (2016).
\newblock {Early Mars serpentinization-derived CH$_{4}$ reservoirs,
  H$_{2}$-induced warming and paleopressure evolution}.
\newblock {\em Meteoritics and Planetary Science}, 51(11):2234--2245.

\bibitem[{Clough} et~al., 2005]{Clough:2005}
{Clough}, S.~A., {Shephard}, M.~W., {Mlawer}, E.~J., {Delamere}, J.~S.,
  {Iacono}, M.~J., {Cady-Pereira}, K., {Boukabara}, S., and {Brown}, P.~D.
  (2005).
\newblock {Atmospheric radiative transfer modeling: a summary of the AER
  codes}.
\newblock {\em Journal of Quantitative Spectroscopy and Radiative Transfer},
  91:233--244.

\bibitem[{Codron}, 2012]{Codron:2012}
{Codron}, F. (2012).
\newblock {Ekman heat transport for slab oceans}.
\newblock {\em Climate Dynamics}, 38:379--389.

\bibitem[{Davis} et~al., 2016]{Davis:2016}
{Davis}, J.~M., {Balme}, M., {Grindrod}, P.~M., {Williams}, R.~M.~E., and
  {Gupta}, S. (2016).
\newblock {Extensive Noachian fluvial systems in Arabia Terra: Implications for
  early Martian climate}.
\newblock {\em Geology}, 44(10):847--850.

\bibitem[{Di Achille} and {Hynek}, 2010]{Diachille:2010}
{Di Achille}, G. and {Hynek}, B.~M. (2010).
\newblock {Ancient ocean on Mars supported by global distribution of deltas and
  valleys}.
\newblock {\em Nature Geoscience}, 3:459--463.

\bibitem[{Dong} et~al., 2018]{Dong:2018mars}
{Dong}, C., {Lee}, Y., {Ma}, Y., {Lingam}, M., {Bougher}, S., {Luhmann}, J.,
  {Curry}, S., {Toth}, G., {Nagy}, A., {Tenishev}, V., {Fang}, X., {Mitchell},
  D., {Brain}, D., and {Jakosky}, B. (2018).
\newblock {Modeling Martian Atmospheric Losses over Time: Implications for
  Exoplanetary Climate Evolution and Habitability}.
\newblock {\em The Astrophysical Journal Letters}, 859(1):L14.

\bibitem[{Fassett} and {Head}, 2008]{Fassett:2008b}
{Fassett}, C.~I. and {Head}, J.~W. (2008).
\newblock {Valley network-fed, open-basin lakes on Mars: Distribution and
  implications for Noachian surface and subsurface hydrology}.
\newblock {\em Icarus}, 198:37--56.

\bibitem[{Fastook} and {Head}, 2015]{Fastook:2015}
{Fastook}, J.~L. and {Head}, J.~W. (2015).
\newblock {Glaciation in the Late Noachian Icy Highlands: Ice accumulation,
  distribution, flow rates, basal melting, and top-down melting rates and
  patterns}.
\newblock {\em Planetary and Space Science}, 106:82--98.

\bibitem[{Forget} et~al., 1999]{Forget:1999}
{Forget}, F., {Hourdin}, F., {Fournier}, R., {Hourdin}, C., {Talagrand}, O.,
  {Collins}, M., {Lewis}, S.~R., {Read}, P.~L., and {Huot}, J.-P. (1999).
\newblock {Improved general circulation models of the Martian atmosphere from
  the surface to above 80 km}.
\newblock {\em Journal of Geophysical Research}, 104:24155--24176.

\bibitem[Forget and Pierrehumbert, 1997]{Forget:1997}
Forget, F. and Pierrehumbert, R.~T. (1997).
\newblock Warming early {Mars} with carbon dioxide clouds that scatter infrared
  radiation.
\newblock {\em Science}, 278:1273--1276.

\bibitem[{Forget} et~al., 2013]{Forget:2013}
{Forget}, F., {Wordsworth}, R., {Millour}, E., {Madeleine}, J.-B., {Kerber},
  L., {Leconte}, J., {Marcq}, E., and {Haberle}, R.~M. (2013).
\newblock {3D modelling of the early martian climate under a denser CO$_{2}$
  atmosphere: Temperatures and CO$_{2}$ ice clouds}.
\newblock {\em Icarus}, 222:81--99.

\bibitem[{Fu} and {Liou}, 1992]{Fu:1992}
{Fu}, Q. and {Liou}, K.~N. (1992).
\newblock {On the correlated k-distribution method for radiative transfer in
  nonhomogeneous atmospheres}.
\newblock {\em Journal of Atmospheric Sciences}, 49:2139--2156.

\bibitem[Godin et~al., 2020]{Godin:2020}
Godin, P., Ramirez, R.~M., Campbell, C., Wizenberg, T., Nguyen, T.~G., Strong,
  K., and Moores, J.~E. (2020).
\newblock Collision-induced absorption of ch4-co2 and h2-co2 complexes and
  their effect on the ancient martian atmosphere.
\newblock {\em Earth and Space Science Open Archive}, page~11.

\bibitem[Gough, 1981]{Gough:1981}
Gough, D.~O. (1981).
\newblock Solar interior structure and luminosity variations.
\newblock {\em Solar Phys.}, 74:21--34.

\bibitem[{Gruszka} and {Borysow}, 1998]{Gruszka:1998}
{Gruszka}, M. and {Borysow}, A. (1998).
\newblock Computer simulation of the far infrared collision induced absorption
  spectra of gaseous co2.
\newblock {\em Molecular Physics}, 93(6):1007--1016.

\bibitem[{Haberle} et~al., 2019]{Haberle:2019}
{Haberle}, R.~M., {Zahnle}, K., {Barlow}, N.~G., and {Steakley}, K.~E. (2019).
\newblock {Impact Degassing of H$_{2}$ on Early Mars and its Effect on the
  Climate System}.
\newblock {\em Geophysical Research Letters}, 46(22):13,355--13,362.

\bibitem[{Halevy} and {Head}, 2014]{Halevy:2014}
{Halevy}, I. and {Head}, III, J.~W. (2014).
\newblock {Episodic warming of early Mars by punctuated volcanism}.
\newblock {\em Nature Geoscience}, 7:865--868.

\bibitem[{Halevy} et~al., 2009]{Halevy:2009}
{Halevy}, I., {Pierrehumbert}, R.~T., and {Schrag}, D.~P. (2009).
\newblock {Radiative transfer in CO$_{2}$-rich paleoatmospheres}.
\newblock {\em Journal of Geophysical Research (Atmospheres)}, 114.

\bibitem[{Haqq-Misra} et~al., 2008]{Haqq-misra:2008}
{Haqq-Misra}, J.~D., {Domagal-Goldman}, S.~D., {Kasting}, P.~J., and {Kasting},
  J.~F. (2008).
\newblock {A Revised, Hazy Methane Greenhouse for the Archean Earth}.
\newblock {\em Astrobiology}, 8:1127--1137.

\bibitem[{Hayworth} et~al., 2020]{Hayworth:2020}
{Hayworth}, B. P.~C., {Kopparapu}, R.~K., {Haqq-Misra}, J., {Batalha}, N.~E.,
  {Payne}, R.~C., {Foley}, B.~J., {Ikwut-Ukwa}, M., and {Kasting}, J.~F.
  (2020).
\newblock {Warming early Mars with climate cycling: The effect of
  CO$_{2}$-H$_{2}$ collision-induced absorption}.
\newblock {\em Icarus}, 345:113770.

\bibitem[{Hourdin} et~al., 2006]{Hourdin:2006}
{Hourdin}, F., {Musat}, I., {Bony}, S., {Braconnot}, P., {Codron}, F.,
  {Dufresne}, J.-L., {Fairhead}, L., {Filiberti}, M.-A., {Friedlingstein}, P.,
  {Grandpeix}, J.-Y., {Krinner}, G., {Levan}, P., {Li}, Z.-X., and {Lott}, F.
  (2006).
\newblock {The LMDZ4 general circulation model: climate performance and
  sensitivity to parametrized physics with emphasis on tropical convection}.
\newblock {\em Climate Dynamics}, 27:787--813.

\bibitem[{Howard}, 2007]{Howard:2007}
{Howard}, A.~D. (2007).
\newblock {Simulating the development of Martian highland landscapes through
  the interaction of impact cratering, fluvial erosion, and variable hydrologic
  forcing}.
\newblock {\em Geomorphology}, 91:332--363.

\bibitem[{Hynek} et~al., 2010]{Hynek:2010}
{Hynek}, B.~M., {Beach}, M., and {Hoke}, M.~R.~T. (2010).
\newblock {Updated global map of Martian valley networks and implications for
  climate and hydrologic processes}.
\newblock {\em Journal of Geophysical Research (Planets)}, 115:E09008.

\bibitem[{Kamada} et~al., 2020]{Kamada:2020}
{Kamada}, A., {Kuroda}, T., {Kasaba}, Y., {Terada}, N., {Nakagawa}, H., and
  {Toriumi}, K. (2020).
\newblock {A coupled atmosphere-hydrosphere global climate model of early Mars:
  A 'cool and wet' scenario for the formation of water channels}.
\newblock {\em Icarus}, 338:113567.

\bibitem[{Karman} et~al., 2019]{Karman:2019}
{Karman}, T., {Gordon}, I.~E., {van der Avoird}, A., {Baranov}, Y.~I.,
  {Boulet}, C., {Drouin}, B.~J., {Groenenboom}, G.~C., {Gustafsson}, M.,
  {Hartmann}, J.-M., {Kurucz}, R.~L., {Rothman}, L.~S., {Sun}, K., {Sung}, K.,
  {Thalman}, R., {Tran}, H., {Wishnow}, E.~H., {Wordsworth}, R., {Vigasin},
  A.~A., {Volkamer}, R., and {van der Zande}, W.~J. (2019).
\newblock {Update of the HITRAN collision-induced absorption section}.
\newblock {\em Icarus}, 328:160--175.

\bibitem[{Kerber} et~al., 2015]{Kerber:2015}
{Kerber}, L., {Forget}, F., and {Wordsworth} (2015).
\newblock {Sulfur in the early martian atmosphere revisited : Experiments with
  a 3-D Global Climate Model}.
\newblock {\em Icarus}, 261:133--148.

\bibitem[{Kite} et~al., 2011]{Kite:2011b}
{Kite}, E.~S., {Rafkin}, S., {Michaels}, T.~I., {Dietrich}, W.~E., and {Manga},
  M. (2011).
\newblock {Chaos terrain, storms, and past climate on Mars}.
\newblock {\em Journal of Geophysical Research}, 116.

\bibitem[{Kite} et~al., 2014]{Kite:2014}
{Kite}, E.~S., {Williams}, J.~P., {Lucas}, A., and {Aharonson}, O. (2014).
\newblock {Low palaeopressure of the martian atmosphere estimated from the size
  distribution of ancient craters}.
\newblock {\em Nature Geoscience}, 7:335--339.

\bibitem[{Kitzmann}, 2016]{Kitzmann:2016}
{Kitzmann}, D. (2016).
\newblock {Revisiting the Scattering Greenhouse Effect of CO$_{2}$ Ice Clouds}.
\newblock {\em The Astrophysical Journal Letters}, 817.

\bibitem[{Laskar} et~al., 2004]{Laskar:2004}
{Laskar}, J., {Correia}, A.~C.~M., {Gastineau}, M., {Joutel}, F., {Levrard},
  B., and {Robutel}, P. (2004).
\newblock {Long term evolution and chaotic diffusion of the insolation
  quantities of Mars}.
\newblock {\em Icarus}, 170:343--364.

\bibitem[{Leconte} et~al., 2013a]{Leconte:2013nat}
{Leconte}, J., {Forget}, F., {Charnay}, B., {Wordsworth}, R., and {Pottier}, A.
  (2013a).
\newblock {Increased insolation threshold for runaway greenhouse processes on
  Earth-like planets}.
\newblock {\em Nature}, 504:268--280.

\bibitem[{Leconte} et~al., 2013b]{Leconte:2013aa}
{Leconte}, J., {Forget}, F., {Charnay}, B., {Wordsworth}, R., {Selsis}, F., and
  {Millour}, E. (2013b).
\newblock {3D climate modeling of close-in land planets: Circulation patterns,
  climate moist bistability and habitability}.
\newblock {\em Astronomy.and Astrophysics, in press}.

\bibitem[{Liggins} et~al., 2020]{Liggins:2020}
{Liggins}, P., {Shorttle}, O., and {Rimmer}, P.~B. (2020).
\newblock {Can volcanism build hydrogen-rich early atmospheres?}
\newblock {\em Earth and Planetary Science Letters}, 550:116546.

\bibitem[{Lillis} et~al., 2017]{Lillis:2017}
{Lillis}, R.~J., {Deighan}, J., {Fox}, J.~L., {Bougher}, S.~W., {Lee}, Y.,
  {Combi}, M.~R., {Cravens}, T.~E., {Rahmati}, A., {Mahaffy}, P.~R., {Benna},
  M., {Elrod}, M.~K., {McFadden}, J.~P., {Ergun}, R.~E., {Andersson}, L.,
  {Fowler}, C.~M., {Jakosky}, B.~M., {Thiemann}, E., {Eparvier}, F., {Halekas},
  J.~S., {Leblanc}, F., and {Chaufray}, J.-Y. (2017).
\newblock {Photochemical escape of oxygen from Mars: First results from MAVEN
  in situ data}.
\newblock {\em Journal of Geophysical Research (Space Physics)},
  122:3815--3836.

\bibitem[{Luo} et~al., 2017]{Luo:2017}
{Luo}, W., {Cang}, X., and {Howard}, A.~D. (2017).
\newblock {New Martian valley network volume estimate consistent with ancient
  ocean and warm and wet climate}.
\newblock {\em Nature Communications}, 8:15766.

\bibitem[{Luo} et~al., 2020]{Luo:2020}
{Luo}, W., {Howard}, A.~D., and {Cang}, X. (2020).
\newblock {Comment on ``The volume of water required to carve the Martian
  valley networks: Improved constraints using updated methods''}.
\newblock {\em Icarus}, 336:113321.

\bibitem[{Malin} and {Edgett}, 2003]{Malin:2003}
{Malin}, M.~C. and {Edgett}, K.~S. (2003).
\newblock {Evidence for Persistent Flow and Aqueous Sedimentation on Early
  Mars}.
\newblock {\em Science}, 302:1931--1934.

\bibitem[{Manabe}, 1969]{Manabe:1969}
{Manabe}, S. (1969).
\newblock {Climate and the Ocean CIRCULATION1}.
\newblock {\em Monthly Weather Review}, 97:739.

\bibitem[{Mangold} and {Ansan}, 2006]{Mangold:2006}
{Mangold}, N. and {Ansan}, V. (2006).
\newblock {Detailed study of an hydrological system of valleys, a delta and
  lakes in the Southwest Thaumasia region, Mars}.
\newblock {\em Icarus}, 180:75--87.

\bibitem[{Matsubara} et~al., 2011]{Matsubara:2011}
{Matsubara}, Y., {Howard}, A.~D., and {Drummond}, S.~A. (2011).
\newblock {Hydrology of early Mars: Lake basins}.
\newblock {\em Journal of Geophysical Research (Planets)}, 116:E04001.

\bibitem[{Menou}, 2013]{Menou:2013}
{Menou}, K. (2013).
\newblock {Water-trapped Worlds}.
\newblock {\em The Astrophysical Journal}, 774.

\bibitem[{Mondelain} et~al., 2021]{Mondelain:2021}
{Mondelain}, D., {Boulet}, C., and {Hartmann}, J.~M. (2021).
\newblock {The binary absorption coefficients for H$_{2}$ + CO$_{2}$ mixtures
  in the 2.12-2.35 {\ensuremath{\mu}}m spectral region determined by CRDS and
  by semi-empirical calculations}.
\newblock {\em Journal of Quantitative Spectroscopy and Radiative Transfer},
  260:107454.

\bibitem[{Moore} et~al., 2003]{Moore:2003}
{Moore}, J.~M., {Howard}, A.~D., {Dietrich}, W.~E., and {Schenk}, P.~M. (2003).
\newblock {Martian Layered Fluvial Deposits: Implications for Noachian Climate
  Scenarios}.
\newblock {\em Geophysical Research Letters}, 30:2292.

\bibitem[{Ozak} et~al., 2016]{Ozak:2016}
{Ozak}, N., {Aharonson}, O., and {Halevy}, I. (2016).
\newblock {Radiative transfer in CO$_{2}$-rich atmospheres: 1. Collisional line
  mixing implies a colder early Mars}.
\newblock {\em Journal of Geophysical Research (Planets)}, 121:965--985.

\bibitem[{Palumbo} et~al., 2018]{Palumbo:2018}
{Palumbo}, A.~M., {Head}, J.~W., and {Wordsworth}, R.~D. (2018).
\newblock {Late Noachian Icy Highlands climate model: Exploring the possibility
  of transient melting and fluvial/lacustrine activity through peak annual and
  seasonal temperatures}.
\newblock {\em Icarus}, 300:261--286.

\bibitem[{Piqueux} and {Christensen}, 2009]{Piqueux:2009}
{Piqueux}, S. and {Christensen}, P.~R. (2009).
\newblock {A model of thermal conductivity for planetary soils: 1. Theory for
  unconsolidated soils}.
\newblock {\em Journal of Geophysical Research (Planets)}, 114:E09005.

\bibitem[{Pollack} et~al., 1987]{Pollack:1987}
{Pollack}, J.~B., {Kasting}, J.~F., {Richardson}, S.~M., and {Poliakoff}, K.
  (1987).
\newblock {The case for a wet, warm climate on early Mars}.
\newblock {\em Icarus}, 71:203--224.

\bibitem[{Poulet} et~al., 2005]{Poulet:2005}
{Poulet}, F., {Bibring}, J.-P., {Mustard}, J.~F., {Gendrin}, A., {Mangold}, N.,
  {Langevin}, Y., {Arvidson}, R.~E., {Gondet}, B., and {Gomez}, C. (2005).
\newblock {Phyllosilicates on Mars and implications for early martian climate}.
\newblock {\em Nature}, 438:623--627.

\bibitem[{Ramirez}, 2017]{Ramirez:2017a}
{Ramirez}, R.~M. (2017).
\newblock {A warmer and wetter solution for early Mars and the challenges with
  transient warming}.
\newblock {\em Icarus}, 297:71--82.

\bibitem[{Ramirez}, 2019]{Ramirez:2019rnaas}
{Ramirez}, R.~M. (2019).
\newblock {Implications of Revised CO$_{2}$-CH$_{4}$ and CO$_{2}$-H$_{2}$
  Absorption for Outer Edge Habitable Zone Planets}.
\newblock {\em Research Notes of the American Astronomical Society}, 3(3):48.

\bibitem[{Ramirez} and {Craddock}, 2018]{Ramirez:2018craddock}
{Ramirez}, R.~M. and {Craddock}, R.~A. (2018).
\newblock {The geological and climatological case for a warmer and wetter early
  Mars}.
\newblock {\em Nature Geoscience}, 11:230--237.

\bibitem[{Ramirez} and {Kasting}, 2017]{Ramirez:2017}
{Ramirez}, R.~M. and {Kasting}, J.~F. (2017).
\newblock {Could cirrus clouds have warmed early Mars?}
\newblock {\em Icarus}, 281:248--261.

\bibitem[{Ramirez} et~al., 2014]{Ramirez:2014}
{Ramirez}, R.~M., {Kopparapu}, R., {Zugger}, M.~E., {Robinson}, T.~D.,
  {Freedman}, R., and {Kasting}, J.~F. (2014).
\newblock {Warming early Mars with CO$_{2}$ and H$_{2}$}.
\newblock {\em Nature Geoscience}, 7:59--63.

\bibitem[{Richard} et~al., 2012]{Richard:2012}
{Richard}, C., {Gordon}, I.~E., {Rothman}, L.~S., {Abel}, M., {Frommhold}, L.,
  {Gustafsson}, M., {Hartmann}, J.-M., {Hermans}, C., {Lafferty}, W.~J.,
  {Orton}, G.~S., {Smith}, K.~M., and {Tran}, H. (2012).
\newblock {New section of the HITRAN database: Collision-induced absorption
  (CIA)}.
\newblock {\em Journal of Quantitative Spectroscopy and Radiative Transfer},
  113:1276--1285.

\bibitem[{Rosenberg} et~al., 2019]{Rosenberg:2019}
{Rosenberg}, E.~N., {Palumbo}, A.~M., {Cassanelli}, J.~P., {Head}, J.~W., and
  {Weiss}, D.~K. (2019).
\newblock {The volume of water required to carve the martian valley networks:
  Improved constraints using updated methods}.
\newblock {\em Icarus}, 317:379--387.

\bibitem[{Rothman} et~al., 2009]{Rothman:2009}
{Rothman}, L.~S., {Gordon}, I.~E., {Barbe}, A., {Benner}, D.~C., {Bernath},
  P.~F., {Birk}, M., {Boudon}, V., {Brown}, L.~R., {Campargue}, A., {Champion},
  J.-P., {Chance}, K., {Coudert}, L.~H., {Dana}, V., {Devi}, V.~M., {Fally},
  S., {Flaud}, J.-M., {Gamache}, R.~R., {Goldman}, A., {Jacquemart}, D.,
  {Kleiner}, I., {Lacome}, N., {Lafferty}, W.~J., {Mandin}, J.-Y., {Massie},
  S.~T., {Mikhailenko}, S.~N., {Miller}, C.~E., {Moazzen-Ahmadi}, N.,
  {Naumenko}, O.~V., {Nikitin}, A.~V., {Orphal}, J., {Perevalov}, V.~I.,
  {Perrin}, A., {Predoi-Cross}, A., {Rinsland}, C.~P., {Rotger}, M., {{\v
  S}ime{\v c}kov{\'a}}, M., {Smith}, M.~A.~H., {Sung}, K., {Tashkun}, S.~A.,
  {Tennyson}, J., {Toth}, R.~A., {Vandaele}, A.~C., and {Vander Auwera}, J.
  (2009).
\newblock {The HITRAN 2008 molecular spectroscopic database}.
\newblock {\em Journal of Quantitative Spectroscopy and Radiative Transfer},
  110:533--572.

\bibitem[{Sagan}, 1977]{Sagan:1977}
{Sagan}, C. (1977).
\newblock {Reducing greenhouses and the temperature history of earth and Mars}.
\newblock {\em Nature}, 269:224--226.

\bibitem[{Schaake} et~al., 1996]{Schaake:1996}
{Schaake}, J.~C., {Koren}, V.~I., {Duan}, Q.-Y., {Mitchell}, K., and {Chen}, F.
  (1996).
\newblock {Simple water balance model for estimating runoff at different
  spatial and temporal scales}.
\newblock {\em Journal of Geophysical Research: Atmospheres}, 101:7461--7475.

\bibitem[{Segura} et~al., 2012]{Segura:2012}
{Segura}, T.~L., {McKay}, C.~P., and {Toon}, O.~B. (2012).
\newblock {An impact-induced, stable, runaway climate on Mars}.
\newblock {\em Icarus}, 220:144--148.

\bibitem[{Segura} et~al., 2008]{Segura:2008}
{Segura}, T.~L., {Toon}, O.~B., and {Colaprete}, A. (2008).
\newblock {Modeling the environmental effects of moderate-sized impacts on
  Mars}.
\newblock {\em Journal of Geophysical Research (Planets)}, 113:E11007.

\bibitem[{Segura} et~al., 2002]{Segura:2002}
{Segura}, T.~L., {Toon}, O.~B., {Colaprete}, A., and {Zahnle}, K. (2002).
\newblock {Environmental Effects of Large Impacts on Mars}.
\newblock {\em Science}, 298:1977--1980.

\bibitem[Smith et~al., 1999]{Smith:1999}
Smith, D.~E., Zuber, M.~T., and {17 coauthors} (1999).
\newblock The global topography of {Mars} and implication for surface
  evolution.
\newblock {\em Science}, 284:1495--1503.

\bibitem[{Smith} et~al., 2001]{Smith:2001}
{Smith}, D.~E., {Zuber}, M.~T., {Frey}, H.~V., {Garvin}, J.~B., {Head}, J.~W.,
  {Muhleman}, D.~O., {Pettengill}, G.~H., {Phillips}, R.~J., {Solomon}, S.~C.,
  {Zwally}, H.~J., {Banerdt}, W.~B., {Duxbury}, T.~C., {Golombek}, M.~P.,
  {Lemoine}, F.~G., {Neumann}, G.~A., {Rowlands}, D.~D., {Aharonson}, O.,
  {Ford}, P.~G., {Ivanov}, A.~B., {Johnson}, C.~L., {McGovern}, P.~J.,
  {Abshire}, J.~B., {Afzal}, R.~S., and {Sun}, X. (2001).
\newblock {Mars Orbiter Laser Altimeter: Experiment summary after the first
  year of global mapping of Mars}.
\newblock 106:23689--23722.

\bibitem[{Steakley} et~al., 2019]{Steakley:2019}
{Steakley}, K., {Murphy}, J., {Kahre}, M., {Haberle}, R., and {Kling}, A.
  (2019).
\newblock {Testing the impact heating hypothesis for early Mars with a 3-D
  global climate model}.
\newblock {\em Icarus}, 330:169--188.

\bibitem[{Tanaka} et~al., 2014]{Tanaka:2014}
{Tanaka}, K.~L., {Robbins}, S.~J., {Fortezzo}, C.~M., {Skinner}, J.~A., and
  {Hare}, T.~M. (2014).
\newblock {The digital global geologic map of Mars: Chronostratigraphic ages,
  topographic and crater morphologic characteristics, and updated resurfacing
  history}.
\newblock {\em Planetary and Space Science}, 95:11--24.

\bibitem[{Tarnas} et~al., 2018]{Tarnas:2018}
{Tarnas}, J.~D., {Mustard}, J.~F., {Sherwood Lollar}, B., {Bramble}, M.~S.,
  {Cannon}, K.~M., {Palumbo}, A.~M., and {Plesa}, A.~C. (2018).
\newblock {Radiolytic H$_{2}$ production on Noachian Mars: Implications for
  habitability and atmospheric warming}.
\newblock {\em Earth and Planetary Science Letters}, 502:133--145.

\bibitem[{Tian} et~al., 2010]{Tian:2010}
{Tian}, F., {Claire}, M.~W., {Haqq-Misra}, J.~D., {Smith}, M., {Crisp}, D.~C.,
  {Catling}, D., {Zahnle}, K., and {Kasting}, J.~F. (2010).
\newblock {Photochemical and climate consequences of sulfur outgassing on early
  Mars}.
\newblock {\em Earth and Planetary Science Letters}, 295:412--418.

\bibitem[Turbet, 2018]{Turbet:2018}
Turbet, M. (2018).
\newblock {\em {Habitability of planets using numerical climate models.
  Application to extrasolar planets and early Mars.}}
\newblock Theses, {Sorbonne Universit{\'e} / Universit{\'e} Pierre et Marie
  Curie - Paris VI}.

\bibitem[{Turbet} et~al., 2020a]{Turbet:2020spectro}
{Turbet}, M., {Boulet}, C., and {Karman}, T. (2020a).
\newblock {Measurements and semi-empirical calculations of CO$_{2}$ + CH$_{4}$
  and CO$_{2}$ + H$_{2}$ collision-induced absorption across a wide range of
  wavelengths and temperatures. Application for the prediction of early Mars
  surface temperature}.
\newblock {\em Icarus}, 346:113762.

\bibitem[{Turbet} and {Forget}, 2019]{Turbet:2019NatSR}
{Turbet}, M. and {Forget}, F. (2019).
\newblock {The paradoxes of the Late Hesperian Mars ocean}.
\newblock {\em Scientific Reports}, 9:5717.

\bibitem[{Turbet} et~al., 2017]{Turbet:2017icarus}
{Turbet}, M., {Forget}, F., {Head}, J.~W., and {Wordsworth}, R. (2017).
\newblock {3D modelling of the climatic impact of outflow channel formation
  events on early Mars}.
\newblock {\em Icarus}, 288:10--36.

\bibitem[{Turbet} et~al., 2020b]{Turbet:2020impact}
{Turbet}, M., {Gillmann}, C., {Forget}, F., {Baudin}, B., {Palumbo}, A.,
  {Head}, J., and {Karatekin}, O. (2020b).
\newblock {The environmental effects of very large bolide impacts on early Mars
  explored with a hierarchy of numerical models}.
\newblock {\em Icarus}, 335:113419.

\bibitem[{Turbet} et~al., 2016]{Turbet:2016}
{Turbet}, M., {Leconte}, J., {Selsis}, F., {Bolmont}, E., {Forget}, F.,
  {Ribas}, I., {Raymond}, S.~N., and {Anglada-Escud{\'e}}, G. (2016).
\newblock {The habitability of Proxima Centauri b. II. Possible climates and
  observability}.
\newblock {\em Astronomy $\&$ Astrophysics}, 596:A112.

\bibitem[Turbet and Tran, 2017]{Turbet:2017jgr}
Turbet, M. and Tran, H. (2017).
\newblock Comment on "radiative transfer in co2-rich atmospheres: 1.
  collisional line mixing implies a colder early mars".
\newblock {\em Journal of Geophysical Research: Planets}, 122(11):2362--2365.
\newblock 2017JE005373.

\bibitem[{Turbet} et~al., 2019]{Turbet:2019spectro}
{Turbet}, M., {Tran}, H., {Pirali}, O., {Forget}, F., {Boulet}, C., and
  {Hartmann}, J.-M. (2019).
\newblock {Far infrared measurements of absorptions by CH$_{4}$ + CO$_{2}$ and
  H$_{2}$ + CO$_{2}$ mixtures and implications for greenhouse warming on early
  Mars}.
\newblock {\em Icarus}, 321:189--199.

\bibitem[{Urata} and {Toon}, 2013]{Urata:2013h2o}
{Urata}, R.~A. and {Toon}, O.~B. (2013).
\newblock {Simulations of the martian hydrologic cycle with a general
  circulation model: Implications for the ancient martian climate}.
\newblock {\em Icarus}, 226:229--250.

\bibitem[{Villanueva} et~al., 2015]{Villanueva:2015}
{Villanueva}, G.~L., {Mumma}, M.~J., {Novak}, R.~E., {Kaufl}, H.~U., {Hartogh},
  P., {Encrenaz}, T., {Tokunaga}, A., {Khayat}, A., and {Smith}, M.~D. (2015).
\newblock {Strong water isotopic anomalies in the martian atmosphere: Probing
  current and ancient reservoirs}.
\newblock {\em Science}, 348:218--221.

\bibitem[{Wang} et~al., 2016]{Wang:2016}
{Wang}, F., {Cheruy}, F., and {Dufresne}, J.-L. (2016).
\newblock {The improvement of soil thermodynamics and its effects on land
  surface meteorology in the IPSL climate model}.
\newblock {\em Geoscientific Model Development}, 9:363--381.

\bibitem[{Wood} et~al., 1992]{Wood:1992}
{Wood}, E.~F., {Lettenmaier}, D.~P., and {Zartarian}, V.~G. (1992).
\newblock {A land-surface hydrology parameterization with subgrid variability
  for general circulation models}.
\newblock {\em Journal of Geophysical Research: Atmospheres}, 97:2717--2728.

\bibitem[{Wordsworth} et~al., 2010]{Wordsworth:2010}
{Wordsworth}, R., {Forget}, F., and {Eymet}, V. (2010).
\newblock {Infrared collision-induced and far-line absorption in dense CO
  $_{2}$ atmospheres}.
\newblock {\em Icarus}, 210:992--997.

\bibitem[{Wordsworth} et~al., 2013]{Wordsworth:2013}
{Wordsworth}, R., {Forget}, F., {Millour}, E., {Head}, J.~W., {Madeleine},
  J.-B., and {Charnay}, B. (2013).
\newblock {Global modelling of the early martian climate under a denser
  CO$_{2}$ atmosphere: Water cycle and ice evolution}.
\newblock {\em Icarus}, 222:1--19.

\bibitem[{Wordsworth} et~al., 2017]{Wordsworth:2017}
{Wordsworth}, R., {Kalugina}, Y., {Lokshtanov}, S., {Vigasin}, A., {Ehlmann},
  B., {Head}, J., {Sanders}, C., and {Wang}, H. (2017).
\newblock {Transient reducing greenhouse warming on early Mars}.
\newblock {\em Geophysical Research Letters}, 44:665--671.

\bibitem[{Wordsworth} et~al., 2021]{Wordsworth:2021}
{Wordsworth}, R., {Knoll}, A.~H., {Hurowitz}, J., {Baum}, M., {Ehlmann}, B.~L.,
  {Head}, J.~W., and {Steakley}, K. (2021).
\newblock {A coupled model of episodic warming, oxidation and geochemical
  transitions on early Mars}.
\newblock {\em arXiv e-prints}, page arXiv:2103.06736.

\bibitem[{Wordsworth}, 2016]{Wordsworth:2016}
{Wordsworth}, R.~D. (2016).
\newblock {The Climate of Early Mars}.
\newblock {\em Annual Review of Earth and Planetary Sciences}, 44:381--408.

\bibitem[{Wordsworth} et~al., 2015]{Wordsworth:2015}
{Wordsworth}, R.~D., {Kerber}, L., {Pierrehumbert}, R.~T., {Forget}, F., and
  {Head}, J.~W. (2015).
\newblock {Comparison of ''warm and wet'' and ''cold and icy'' scenarios for
  early Mars in a 3-D climate model}.
\newblock {\em Journal of Geophysical Research (Planets)}, 120:1201--1219.

\end{thebibliography}
\bibliographystyle{apalike}


\end{document}